\documentclass[preprint,12pt]{elsarticle}

\usepackage{graphicx}

\usepackage{amssymb}

\journal{Physics Reports}

\begin{document}

\begin{frontmatter}



\title{The Search for a Primordial Magnetic Field}


\author{Dai G. Yamazaki}
\address{National Astronomical Observatory of Japan, Mitaka, Tokyo 181-8588, Japan}
\author{Toshitaka Kajino}%
\address{National Astronomical Observatory of Japan, Mitaka, Tokyo 181-8588, Japan}
\address{Department of Astronomy, Graduate School of Science, The University of Tokyo,
Hongo 7-3-1, Bunkyo-ku, Tokyo 113-0033, Japan}
\author{Grant J. Mathews}%
\address{Department of Physics, Center for Astrophysics, University of Notre Dame, Notre Dame, IN 46556, U.S.A.}
\author{Kiyotomo Ichiki}%
\address{Department of Physics and Astrophysics, Nagoya University, Nagoya 464-8602, Japan}

\begin{abstract}
Magnetic fields appear wherever plasma and currents can be found.  As such, they thread through all scales in Nature.  It is natural, therefore, to suppose that magnetic fields might have been  formed within  the high temperature environments of the big bang.   Such a primordial magnetic field (PMF)  would be expected to arise from and/or influence a variety of cosmological phenomena such as  inflation,  cosmic phase transitions, big bang nucleosynthesis, the cosmic microwave background (CMB) temperature and  polarization anisotropies, the cosmic gravity wave background, and  the formation of large-scale structure.  In this review, we summarize  the development of theoretical models for analyzing the observational consequences  of a PMF.  We also summarize the current state of the art in the search for observational evidence of a PMF.  In particular we review  the framework needed 
to calculate the effects of a PMF power spectrum on the CMB and the development of large scale structure.  We  summarize the current  constraints on the PMF  amplitude $B_\lambda$ and the power spectral index $n_B$  and  discuss prospects for better determining these quantities in the near future.

\end{abstract}


\end{frontmatter}


 \newcounter{one}
 \setcounter{one}{1}
 \newcounter{two}
 \setcounter{two}{2}
 \newcounter{three}
 \setcounter{three}{3}
\section{\label{s1}Introduction}
It is well known that  magnetic fields influence  many physical processes over a broad range of scales in the universe. 
 It is natural, therefore, to suppose that currents and associated magnetic fields could arise from the flow of material in the  high temperature environment of the big bang. Such a primordial magnetic field (PMF) would be expected to manifest itself in the cosmic microwave background (CMB) temperature and polarization anisotropies
\cite{Subramanian:1998fn,Mack:2001gc,Subramanian:2002nh,Lewis:2004ef,Yamazaki:2004vq,Kahniashvili:2005xe,Challinor:2005ye,Dolgov:2005ti,Gopal:2005sg,Yamazaki:2005yd,Kahniashvili:2006hy,Yamazaki:2006bq,Yamazaki:2006ah,Giovannini:2006kc,Yamazaki:2007oc,Paoletti:2008ck,Yamazaki:2008bb,2008nuco.confE.239Y,Sethi:2008eq,Kojima:2008rf,2008PhRvD..78f3012K,Giovannini:2008aa,2010PhRvD..81b3008Y,2010PhRvD..81j3519Y,2010AdAst2010E..80Y}, and also in the formation of large-scale structure (LSS)
\cite{Sethi:2003vp,Sethi:2004pe,Yamazaki:2006mi}.
These  studies  have pointed out that the effects of a PMF are not negligible if the PMF  had a field strength $B_\lambda \sim 1.0$ nG on $\sim 1$ Mpc scales at the epoch of photon last scattering, $z \sim 1100$.
Therefore, it is important to constrain the PMF parameters by cosmological observations.  The purpose of this review is to summarize progress in the development of the calculational techniques to deduce such constraints from existing observational data.  First, however, we motivate this endeavor with a brief discussion of why one thinks there might be a PMF from both theoretical grounds and observations of galaxy clusters.

\subsection{\label{s1_1}Generation models of a PMF}
Many authors have been actively studying  the origin of cosmological primordial magnetic fields.  Possible generation mechanism have recently been reviewed in detail by \cite{2011PhR...505....1K}.  In brief, 
such  primordial fields are expected to have  random distribution of orientations and field strengths.
If the PMF has a nearly scale invariant spectrum like the primary scalar spectrum, one of the best models is an origin from vector potentials generated during the inflation epoch \cite{Turner:1987bw,Ratra:1991bn,Bamba:2004cu}. 
Several authors have proposed that a PMF with a bluer spectrum could have been produced during one of the subsequent cosmological phase transitions \cite{Vachaspati:1991nm,Kibble:1995aa,Ahonen:1997wh,Joyce:1997uy}.
A magnetic field could also have been generated during or after the epoch of photon last scattering  ($z ^<_\sim  1100$) by   vorticities in the cosmological fluid
\cite{Takahashi:2005nd,Hanayama:2005hd,ichiki:2006sc}.
Since each model for the generation of the PMF involves different length scales, the spectral index of the PMF power spectrum, $n_\mathrm{B}$, will depend upon which is the correct  model for the generation of the PMF. Therefore, constraining $n_\mathrm{B}$ phenomenologically can lead to constraints on the epoch for the  generation of  the PMF.  
\subsection{\label{s1_2}Observational evidence for a PMF}
Observations of synchrotron emission \cite{2004IJMPD..13.1549G} and Faraday rotation \cite{Wolfe:1992ab,Clarke:2000bz,Xu:2005rb} in galaxy clusters all indicate the presence of magnetic fields on large scales.  Moreover,  magnetic fields with a strength of $B \sim 1.0$ $\mu$G have been detected in several  galaxy clusters \cite{Wolfe:1992ab,Clarke:2000bz,Xu:2005rb}.   Simulations of such cluster  magnetic fields  may require\cite{2011arXiv1112.0340B} the existence of a primordial  seed magnetic field.  

Since the diffusion time of a magnetic field in galaxy clusters is much longer than the age of the universe, such a magnetic field is "frozen-in" to the ionized fluids \cite{Mack:2001gc}. 
The energy density of the baryon fluid scales as $\rho_b \propto a^{-3}$ while
the magnetic energy density  scales as $\rho_\mathrm{B} \propto a^{-4}$. Therefore, one can relate the strength of the magnetic field energy density to the energy density of the cosmic baryon fluid $B^3 \propto \rho_b^2$.
If clusters of galaxies collapse nearly isotropically relative to the background space, an observed magnetic field of $B \sim 1.0~\mu $G in galaxy clusters now  corresponds to a PMF of order $\sim 1.0$ nG at the epoch of photon last scattering near $z \sim 1100$. 
\subsection{\label{s1_3}Constraint on a PMF by Cosmological observations}
As noted above  there is  a large difference between  the strength of any cosmological magnetic field during  the generation epoch and that which could be detected at the present time. In most currently proposed theoretical scenarios, the generation of a PMF  involves some physical process that generates current  in the early universe {\it before} the epoch of photon last scattering.
However, cosmological observation determine the CMB anisotropies and the matter power spectrum on various cosmological scales {\it during} and {\it after} the epoch of photon last scattering. It is therefore timely and desirable to construct precise theoretical models with which  to analyze the CMB anisotropies and the large-scale-structure (LSS) matter power spectrum as obtained in  present and soon to be obtained  cosmological observations. The purpose of this review is to summarize the development of such models.
\section{\label{s2}Models for the Generation of a PMF}
The origin of  a cosmological primordial magnetic field has been an area of active research by many authors [See \cite{2011PhR...505....1K} for an excellent review].  The proposed models can be divided into three classes which we  briefly summarize.
First, however, note that the cosmological magnetic field damps as 
\begin{eqnarray}
B_0 \propto a^{-2},
\end{eqnarray}
where $a$ is the cosmological scale factor. 
In order to avoid confusion, therefore, it is best  refer to a comoving PMF field strength $B_0$ scaled to the present-day value on some length scale (taken here to be 1 Mpc), i.e. we write 
\begin{eqnarray}
B_{0}=B(\tau)\times
	\left(
		\frac{a(\tau)}{a_{0}(\tau)}
	\right)^2,
\end{eqnarray}
where $\tau$ is the conformal time defined by the differential $d\tau = adt$.
\subsection{\label{s2_1}Inflation}
During the epoch of inflation small quantum perturbations are believed to have been  extended and enlarged  into those of super horizon scales \cite{Liddle:2000booka}.
Since spacetime in  normal inflation is conformally flat \cite{Liddle:2000booka}, the  spacetime and any associated  electromagnetic field are independent and invariant to a conformal transformation. Therefore, unless this invariance to a conformal transformation is broken, a magnetic field cannot be generated by the expanding spacetime \cite{Turner:1987bw,Ratra:1991bn,Bamba:2004cu}. 

In \cite{Turner:1987bw}, however,  it was 
assumed that the electromagnetic field and the gravitational field interacted in such a way as to break the invariance of the electromagnetic fields to a conformal transformation.  They then showed that a magnetic field could  be generated by quantum perturbations.
There are several means to  naturally break such invariance. 
For example, the strength of a PMF generated by an electromagnetic tensor of the dilaton type, has been  estimated to be $\sim 1$nG and $\sim 10^{-5}$nG by \cite{Bamba:2004cu}
and 
\cite{Lemoine:1995dm}, 
respectively. 
\subsection{\label{s2_2}EW transition/QCD transition}
Cosmological phase transitions could also produce a PMF with a bluer spectrum \cite{Vachaspati:1991nm,Kibble:1995aa,Ahonen:1997wh,Joyce:1997uy} than that of inflation models.
For example, bubbles of a new  lower temperature phase can be made at the cosmological quark-hadron \cite{Quashnock:1989jm} or electroweak \cite{1996PhRvD..53..662B} phase transitions. As those bubbles collide and percolate, the baryon symmetry is broken leading to the generation of a magnetic field 
\cite{Quashnock:1989jm}.
The strength of the magnetic field generated at these epochs has been estimated  to  be  $\sim 10^{-7}$nG for fields generated by the quark-hadron transition \cite {Quashnock:1989jm} and $\sim 10^{-14} - 10^{-8}$nG for the electroweak transition \cite {1996PhRvD..53..662B}.
\subsection{\label{s2_3}Decoupling and afterward}
Magnetic field generation can also occur on smaller scales during or after the epoch of photon last scattering  $(z < 1000)$  \cite {Hanayama:2005hd,Takahashi:2005nd,ichiki:2006sc}.
If there were eddies before the recombination epoch, they could generate a PMF \cite{ichiki:2006sc}. 
Even without turbulent eddies, the known CMB temperature fluctuations imply the generation of at least some magnetic primordial field.  Since  protons have  mass and photons do not,
photons scatter electrons differently than  protons.  This difference induces 
electric currents as they fall in and out of gravitational potentials. These  electric fields and can generate a magnetic field of about $10^{-9}$nG at 1 Mpc.
\cite{ichiki:2006sc}.
\section{\label{s3}Constraints on the Physics of a PMF}
In this section we briefly introduce that the effects of a PMF on
 Big-Bang Nucleosynthesis (BBN),
the Gravitational Wave Background (GWB),
the Cosmic Microwave Background (CMB) and
Large Scale Structure (LSS). We then
 qualitatively discuss constraints on the PMF parameters deduced 
from each observation.
Following that we shall quantitatively review how the PMF parameters are constrained 
by multiple cosmological observations.

Of course one does not really know {\it a priori}  the spectrum of cosmological magnetic fields.  As a starting point, however, it is natural to presume that one can  characterize the PMF parameters  in the same way one does for photons, i.e. by a power-law spectrum of PMF energy density fluctuations on different wave numbers $k$, i.e. 
\begin{equation}
P(B_k) \propto k^{n_B } ~~,
\end{equation}
where $P(B_k)$ denotes the power in magnetic field energy density on the scale of wave number $k$ and $n_B$ is a spectral index.

In previous work \{e.g. Refs.\cite{Yamazaki:2004vq,Yamazaki:2006bq}\} approximate power law
spectra were utilized to describe a PMF. These approximate spectra have
been very convenient for exploring the qualitative effects of a PMF on various
physical phenomena. However, in this approximation one 
cannot obtain quantitatively accurate results. For example, one could not accurately
constrain the parameters of a PMF from cosmological observations.
In Yamazaki et al. (2008)\cite{Yamazaki:2007oc}, however, a power law formulation for a PMF was
developed without approximation (See Appendix A for details). Using this
formula, we have been able to calculate the effects of a PMF by accounting
for the time evolution of the cut off scale to high accuracy. Also, we
have constructed a numerical program without approximation, for studying the
effects of a PMF on the cosmological perturbations.

Since our main purpose is to review how parameters of the PMF are constrained by  cosmological observations, we first make  a brief  explanation of  the various ways in which a PMF affects  cosmology. The reader who wishes to understand PMF effects in more detail is referred to the following papers regarding 
the CMB: Ref.\cite{Mack:2001gc,Yamazaki:2004vq,Yamazaki:2006bq,Yamazaki:2007oc,2010PhRvD..81b3008Y},
 LSS: Ref.\cite{Sethi:2003vp,Sethi:2004pe,Yamazaki:2006mi,Yamazaki:2008bb,2010PhRvD..81j3519Y}, and
the GWB: Ref.\cite{Caprini:2001nb}.
After reviewing the observational constraints  we will  conclude  with a discussion of the constraints upon possible origins of a PMF based on the present constraints on the PMF parameters from fits to the available observations.
\subsection{\label{s3_1}Constraints from BBN and the GWB}
The balance between the expansion rate of the universe and various  particle reaction rates has important effects on the nucleosynthesis of  light elements in the big-bang.
Moreover, since the energy density of the GWB, $\rho_{\mathrm{GW}}$,
contributes to the total energy density of the universe,
the expansion rate is affected by the GWB.
Therefore, we can indirectly constrain the energy density  $\rho_{\mathrm{GW}}$ from the  light element primordial abundances inferred from observations  of deuterium (D), $^3$He, $^4$He, and $^7$Li.  Following  \cite{2000PhR...331..283M} one can relate the upper limit to the energy density in gravity waves to the upper limit on the effective number of neutrino flavors $N_\nu$ present during BBN:
\begin{eqnarray}
\int^\infty_0
	d\log{(\nu)}
	h^2_0
	\Omega_\mathrm{GWB}(\nu)
\le
5.6\times 10^{-6}
(N_\nu -3) ~~,
\label{eq:gw_Nn}
\end{eqnarray}
where $\nu$ is the frequency of the GWB, $h_0$ is the Hubble parameter (in units of 100 km s$^{-1}$ Mpc$^{-1}$), 
$\Omega_\mathrm{GWB}$ is the ratio of the energy density of the GWB $\rho_\mathrm{GWB}$ to 
the critical density of the universe $\rho_\mathrm{c}$.

The upper limit to the number of neutrino flavors $N_\nu$ is constrained from a variety of observations.  If one only utilizes a  comparison between BBN and the inferred primordial abundances of deuterium (D) and helium ($^4$He) in the standard big bang one obtains  \cite{2005APh....23..313C}:
\begin{eqnarray}
N_\nu &=& 3.14^{+0.70}_{-0.65}~\mathrm{at~68\%~CL} \nonumber \\
      & & (2.49 \le N_\nu \le 3.84 ~\mathrm{at~68\%~CL})~~.
\end{eqnarray}
Alternatively, if one considers constraints from the WMAP \cite{2007ApJS..170..288H,2007ApJS..170..335P,2007ApJS..170..377S,2009ApJS..180..330K} CMB power spectrum analysis including Luminous Red Galaxies (LRG) \cite{Ross:2008ze,Rozo:2007yt} and the Hubble constant, $H_0$ measurements,  this limit increases to \cite{2011ApJS..192...18K}
\begin{eqnarray}
3.45 \le N_\nu \le 5.01 ~\mathrm{at~68\%~CL}~~.
\end{eqnarray}
Or, if one also  includes baryon acoustic oscillations (BAO) in the analysis this limit becomes \cite{2011ApJS..192...18K}
\begin{eqnarray}
3.46 \le N_\nu \le 5.20 ~\mathrm{at~68\%~CL}~~.
\end{eqnarray}
An analysis that also includes constraints from Type Ia supernovae (SN)  $WMAP+BAO+SN+HST$
leads to \cite{2011ApJS..192...18K}
\begin{eqnarray}
2.90 \le N_\nu \le 5.90 ~\mathrm{at~68\%~CL}~~.
\end{eqnarray}
Therefore, depending upon which constraint is adopted, one can rewrite Eq.(\ref{eq:gw_Nn}) as:
\begin{eqnarray}
\int^\infty_0
	d\log{(\nu)}
	h^2_0
	\Omega_\mathrm{GWB}(\nu)
&&\le
4.7\times 10^{-6} \mathrm{BBN,~D~and~^4He},\\
\le&&
1.13\times 10^{-5} (WMAP + LRG + H_0),\\
\le&&
1.23\times 10^{-5} (WMAP + BAO + H_0),\\
\le&&
1.62\times 10^{-5} (WMAP+BAO+SN+HST). \nonumber \\
\label{eq:gw_Nn}
\end{eqnarray}

\begin{figure*}[h]
\includegraphics[width=1.0\textwidth]{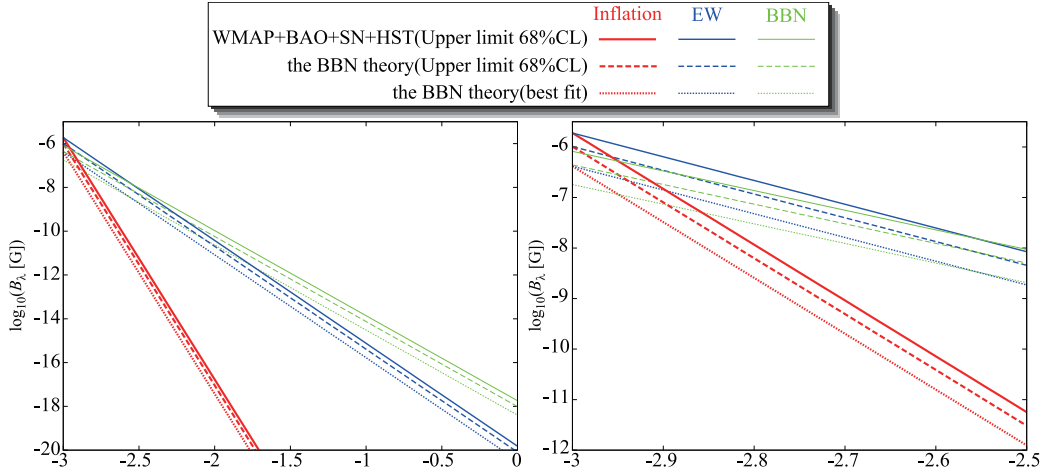}
\caption{\label{gw} 
Constrained parameter plane of the PMF field strength $B_\lambda$ vs. the power spectral index $n_\mathrm{B}$ by the BBN.
The thin (green), middle (blue), and bold (red) lines show upper limits on the produced PMF during the epoch of BBN, the electroweak transition, and the inflation epoch, respectively.
 (For interpretation of the references to colour in this figure legend, the reader is referred to the web version of this article.)
}
\end{figure*}

Caprini and Durrer  \cite{Caprini:2001nb} have analyzed the generation of a GWB from fluctuations in the PMF.  They then used the BBN constraints on the GWB to place surprisingly  strong constraints on the parameters characterizing a PMF.
Following  \cite{Caprini:2001nb}, we have recalculated the GWB upper limits  on  the PMF field strength as a function of  spectral index for a PMF formed during different epochs.  These are shown\footnote{Caprini and Durer (2004) quoted field strength  for a comoving scale of galaxies $\lambda = 0.1$  Mpc, however our results are quoted for galaxy cluster scales $\lambda = 1$ Mpc. } in  Fig.\ref{gw}. The right part of this figure is an expanded plot for spectral indices $n_B$ from -3 to -2.5.
The green, blue, and red lines on this plot are  the upper limits to a  PMF generated during  big-bang nucleosynthesis,  the electroweak transition, or  the inflation epoch, respectively.
For this plot we adopt the WMAP constraint on $N_\nu$ from the $WMAP+BAO+SN+HST$ analysis \cite{2011ApJS..192...18K}.
 
If the PMF were generated during  inflation, the power should be  nearly scale invariant ($n_B \sim -3$) \cite{Mack:2001gc}. 
In this case, the PMF parameters are not constrained strongly by the associated  GWB.
For example,  PMF fields strengths as large as  $B_\lambda \sim 10^{-7}$G are allowed for a spectral index  of $n_\mathrm{B}\sim -3$ (right panel of Fig.\ref{gw}). 
As we review  later, the constraint on the PMF strength from various  cosmological observations is less than several nano Gauss.
Therefore, if the PMF were generated during the inflation epoch and the PMF  power spectrum is  nearly scale-invariant, the PMF strength is not constrained very much by this method.

On the other hand, if the PMF were  generated after the inflation epoch it has a bluer spectrum (smaller spectral index) and is better constrained by the GWB. 
One can deduce from  the left panel of Fig.\ref{gw}, that the strength of a PMF generated during  the EW epoch has an upper limit  of order $10^{-11}$ G, $10^{-15}$ G or $10^{-20}$ G for  $n_\mathrm{B} = $ -2, -1 or 0, respectively. Similarly,  if the PMF were  generated during  the BBN epoch, the upper limit of the PMF strength  is constrained to be order $10^{-10}$ G, $10^{-14}$ G or  $10^{-18}$ G for  $n_\mathrm{B} = $ -2, -1 or 0, respectively.

Furthermore, since the upper limit on the PMF amplitude, $B^\mathrm{up}_\lambda$, depends upon the effective number of neutrinos $N_\nu$  allowed in the early universe:
\begin{eqnarray}
B^\mathrm{up}_\lambda \propto \sqrt{N_\nu-3},
\end{eqnarray}
the constraints on the PMF parameters strongly depend upon independent observations which constrain $N_\nu$.

There has also been an attempt to constrain the PMF field strength  from direct measurements of limits to the present cosmological GWB. \cite{2010PhRvD..81b3002W} has deduced the constraint $B_\lambda < 4 \times 10^{-7}$G at 1 Mpc and $B_\lambda < 9 \times 10^{-11}$G at 100 Mpc from the LIGO S5 data
\cite{LIGO_S5}.
\subsection{\label{s3_2}Constraints from the CMB}
For a PMF of order 1.0-10 nG at the surface of photon last scattering, the total energy density in the PMF is much smaller than that of the temperature fluctuations of the CMB (called the $T$-mode, where $T$ denotes the total scalar temperature).  Therefore, we can treat the energy density of a PMF as a first order perturbation and assume a stiff source for the time evolution \cite{Mack:2001gc}. That is, all back reactions from the fluid onto the magnetic field can be discarded because they are second order perturbations. In this case, we can also assume that the conductivity of the primordial plasma is very large and that the electric field is negligible, i.e. $E\sim 0$. This ''frozen-in'' condition is a very good and useful approximation \cite{Mack:2001gc}.  

Recently, the CMB polarization power spectra have been observed by several projects. This has allowed for more precise constraints \cite{2010PhRvD..81b3008Y,2010PhRvD..82h3005K,2011PhRvD..83l3533P} on  the PMF and other cosmological parameters. 
For clarity, we first  give a brief description of the CMB polarization as follows:
A photon scattered by an electron is polarized perpendicular to the incident direction.
When the incident photons are isotropic or have only a dipole distribution, there is no net polarization of the scattered photons. On the other hand, if  the incident photons are perpendicular to each other and have different intensities, the scattered photons will have a net linear polarization. 
In addition,  photons are polarized by the perturbed gravitational potential, e.g. weak lensing effects. 
The polarization of  photons with  positive parity is   called the "$E$-mode", 
and the  polarization  with negative parity is called the "$B$-mode".
Therefore,  there are three observable modes: the  $T$-mode; the  $E$-mode and the $B$-mode.
From these, one can generate correlations leading to three power spectra denoted: $TT$, $EE$, and $BB$.  There are also three cross-correlation spectra: $TE$, $TB$, and $EB$. 
The only nonzero spectra, however, are the $TT$, $EE$, $BB$, and $TE$ modes due to parity considerations. Also,  the modes can have up to  three kinds of fluctuations: scalar, vector and tensor.

\begin{figure*}[h]
\includegraphics[width=1.0\textwidth]{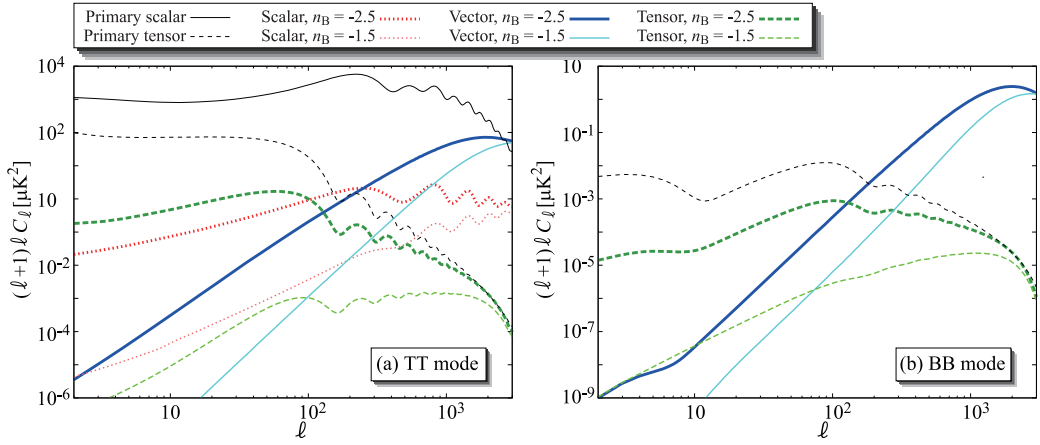}
\caption{\label{TTBB} 
Temperature fluctuations [$TT$ mode on Panel (a)] and polarization anisotropies [$BB$ mode on Panel (a)] of the CMB from the primary energy density and the PMF. We plot models with $B_\lambda = 3.0$ nG and $n_\mathrm{B} = -1.5$ and -2.5 as labeled. The scalar to tensor ratio for this plot is taken to be 0.2 for the primary tensor contribution.
}
\end{figure*}

Figure \ref{TTBB} illustrates  that  a PMF produces the largest  temperature fluctuations and polarization anisotropies of the CMB for higher multipole moments $\ell$.
 The energy density of the PMF is proportional to the fourth power of the scale factor $a^{-4}$ just like the photon energy density. On the other hand, the field strength of the PMF scales with the plasma density $\rho_{p0}$ as 
\begin{eqnarray}
B_\lambda (k,a) = B_\lambda (k)*(\rho_{p0} + \delta \rho_p)
\label{eq:B_rho}
\end{eqnarray}
where $\delta \rho_p$ is a first order perturbation in the plasma density. 
If the field strength of the PMF is less than the order of 10 nG, the energy density of the PMF is  less than or equal to the perturbations in  the energy density of the CMB photons. 
Thus, the second term of Eq.(\ref{eq:B_rho}) is a second order perturbation and negligible in the linearized  theory.
Therefore, we can ignore the effect of the time evolution of the plasma density fluctuations (e.g. Silk damping \cite{1968ApJ...151..459S}) on the PMF in the linear theory.
As a result, the effects of a PMF on the temperature fluctuations and polarization anisotropies of the CMB tend to be large even  for scales smaller than the Silk damping scale.

For the vector mode, the CMB polarization due to a  PMF is largest for higher multipole moments, $\ell$, while the tensor mode of the PMF diminishes for large multipoles [cf. Fig.\ref{TTBB}]  \cite{Seshadri:2000ky,Lewis:2004ef,Yamazaki:2007oc,2010PhRvD..81b3008Y}.
The reason for this is that the gravity waves from the PMF can be negligibly small after horizon crossing. They are small because, once inside the horizon,  the homogeneous solution for the gravity waves begins to oscillate  and decay rapidly.
\cite{1979ZhPmR..30..719S,1982PhLB..115..189R,1985SvA....29..607P,Pritchard:2004qp}. 
As a result, the effect of  gravity waves from the PMF  only occurs  on 
scales larger than the horizon at the epoch of the generation of the CMB ($z\sim 1000$).

\begin{figure*}[h]
\includegraphics[width=1.0\textwidth]{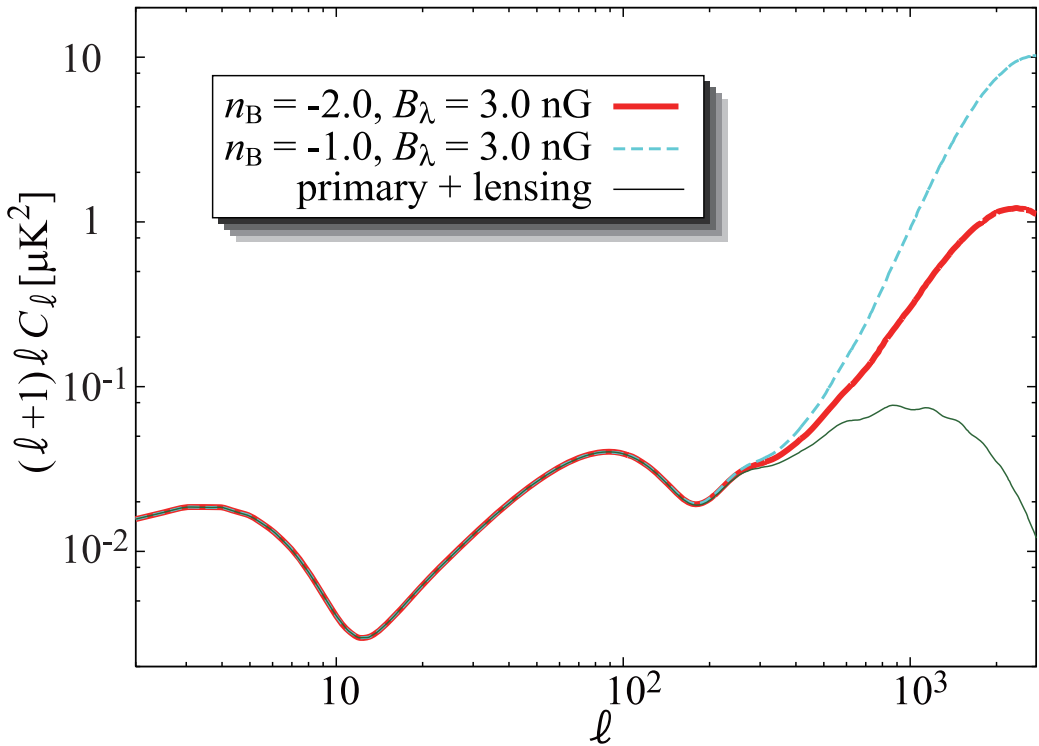}
\caption{\label{BB} 
BB mode from the PMF.
The bold (red) and dashed (azure) curves show results for $(B_\lambda, n_\mathrm{B}) = (3.0 \mathrm{nG}, -2.0)$ and $(B_\lambda, n_\mathrm{B}) = (3.0 \mathrm{nG}, -1.0)$. The thin (green) curve shows the primary tensor mode power spectrum (without the PMF) with weak lensing effects included.
 (For interpretation of the references to colour in this figure legend, the reader is referred to the web version of this article.)
}
\end{figure*}

There are several degeneracies between the PMF and other physical processes in the early universe.
Panel (b) on Figures \ref{TTBB} and \ref{BB} shows 
the BB mode of the CMB polarization from the primary fluctuations,  gravitational lensing, and the PMF.
We can see that the total BB mode spectrum is dominated by the PMF for $B_{\lambda}\gtrsim 2.0$ nG and $\ell \gtrsim 200$. 
We must, however, consider the degeneracy between the PMF and lensing effects on such small angular scales. This is  because the BB mode is converted from the EE mode by a gravitational lensing effect on these angular scales \cite{Pritchard:2004qp}. 
However, the source of the spectrum from the gravitational lensing signal is the $EE$-mode of the CMB polarization of the primary scalar fluctuations. Fortunately, however, this spectrum can be subtracted directly because the $EE$-mode polarization of the primary scalar fluctuations  have been determined independently.


Since the Sunyaev-Zel'dovich effect is one of the main foreground sources for higher $\ell$
\cite{1980ARA&A..18..537S}, we must also  consider the degeneracy between the PMF and the Sunyaev-Zel'dovich effect on the temperature fluctuations of the CMB.
Nevertheless, 
the effects of a PMF on the CMB  background are independent of frequency and the Sunyaev-Zel'dovich effect on the CMB as a foreground depends upon the frequency.  Therefore,  it should be possible  to distinguish the effects of a PMF from such foreground effects by making observations at different frequencies.

The  qualitative features of the constrained parameters of the PMF from only the CMB observations are easy to understand. 
Table \ref{CMBonlyTable} shows the constraint on the PMF parameters from  the combined CMB observations of WMAP, ACBAR and CBI using Markov Chain Monte Carlo (MCMC) methods \cite{cosmomc}.
We find that there is  no obvious degeneracy between the PMF and cosmological parameters. The reason for this is that the effects of a PMF dominate for $\ell > 1000$, while the other cosmological parameters are constrained by the WMAP power spectrum obtained for $\ell < 1000$. 

\begin{figure*}[h]
\includegraphics[width=1.0\textwidth]{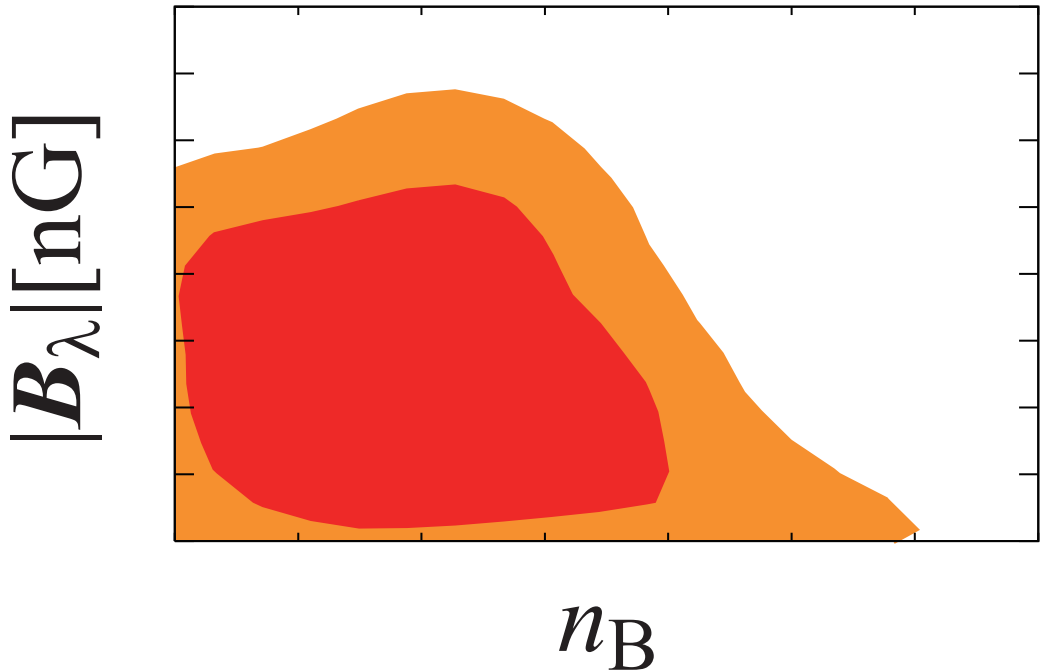}
\caption{\label{PMFofP}
Probability contour plane of $n_\mathrm{B}$ vs. $B_\lambda$ from an MCMC analysis of the CMB and LSS observations. 
Deep (red) and pale (orange) colour contours show the 1 $\sigma$(68\%)
 and 2 $\sigma$(95\%) confidence limits. 
 (For interpretation of the references to colour in this figure legend, the reader is referred to the web version of this article.)
}
\end{figure*}

On the other hand, from Fig. \ref{PMFofP} one can see that there is a strong degeneracy between the PMF amplitude $B_\lambda$ and the power law spectral index $n_\mathrm{B}$.
There are two main reasons for this. 
For one, the effect of a PMF on the CMB is to produce  a peak in the  range of $\ell \ge 1500$.  Unfortunately, however,
observations  are not yet precise enough in this multipole range to constrain the PMF parameters.
This is especially true for the BB mode for higher $\ell$ where we do not yet have  observations of the BB mode with sufficiently small error.
The second reason is simply that the PMF parameters themselves induce similar effects on  the amplitude of the CMB power spectrum \cite{Yamazaki:2006bq,Yamazaki:2007oc}.
\subsection{\label{s3_3}Constraints from LSS}
There are strong constraints on parameters of the PMF from the observed limits on the $\sigma_8$ parameter [see \cite{Yamazaki:2006mi} for details]. This parameter  is determined from  a weighted integral of the matter power spectrum \cite{Peebles:1980booka} and corresponds to the root-mean-square of the matter density fluctuations in a comoving sphere of radius $8h^{-1}$ Mpc.

\begin{figure*}[h]
\includegraphics[width=1.0\textwidth]{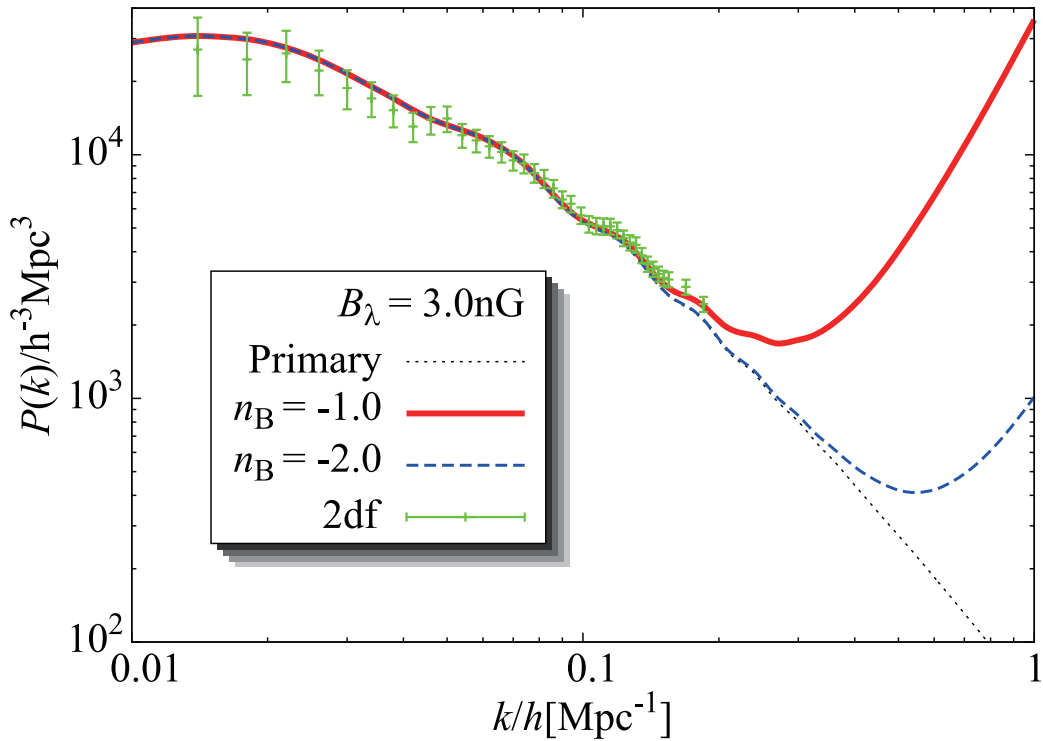}
\caption{\label{mp_nb-10L}
Matter power spectra of the primary CMB, the observational result, and the PMF for   $B_\lambda = 3.0$ nG.
Bold (red) and dashed (blue) curves are for $n_\mathrm{B} = -2.0$ and $-1.0$.
The dotted (black) curve is the primary CMB power spectrum. The dots with error bars are from an analysis of the 2dF data\cite{Cole:2005sx}. 
 (For interpretation of the references to colour in this figure legend, the reader is referred to the web version of this article.)
}
\end{figure*}

Figure.~\ref{mp_nb-10L} shows that the density fluctuations of matter are more strongly affected by a PMF for wavenumbers $k/h >  0.1$ Mpc$^{-1}$\cite{Yamazaki:2006mi}. 
As mentioned  in Sec. \ref{s3_2}, 
 the time evolution of the PMF energy density does not depend on the time evolution of the plasma density fluctuations (e.g. Silk damping) in the linear theory.  Therefore, 
the PMF can survive as a source of  temperature fluctuations and polarization anisotropies of the CMB on scales well below the photon diffusion length.

The baryons influence the cold dark matter(CDM) through gravity.
This effect is very small before the epoch of photon last scattering ($z\simeq 1000$) because the baryon density oscillates with  the photons and their gravitational effect on the CDM becomes very small.
However, after the baryons decouple from the photons, 
the baryon density begins  to affect the density of the CDM through gravitational interactions \cite{Yamamoto:1997qc}.
Also,  the baryons are influenced by any PMF present in the early universe. Thus, a PMF can indirectly affect the CDM evolution. 
Nevertheless, it is reasonable to  assume that the energy density of the PMF is independent of the matter density fluctuations.
In this case, the PMF merely increases the matter power spectrum independently of whether the  pressure or tension dominate the PMF \cite{Yamazaki:2006mi}.

As discussed in  Sec.~\ref{s3_2}, 
there is the strong degeneracy between the PMF amplitude $B_\lambda$ and the power law spectral index $n_\mathrm{B}$. 
Thus, one  needs a good  Bayesian prior constraint  to effectively decouple these two parameters characterizing  the PMF.  
In this regards one is  aided by the observed constraints on the $\sigma_8$ parameter.  The parameter  $\sigma_8$ is constrained by observational data on linear cosmological scales \cite{Cole:2005sx,Tegmark:2006az,Rozo:2007yt,Ross:2008ze} to lie in the range $0.7 < \sigma_8 < 0.9$. 

Fortunately, the recent CMB observations have determined  cosmological parameters based upon the power spectrum obtained on larger scales ($\ell < 1000$)
\cite{2007ApJS..170..377S,2007ApJS..170..288H,2007ApJS..170..335P}. 
The effects of the PMF are mainly  on smaller scales ( $\ell > 1000$)\cite{Yamazaki:2006bq,Yamazaki:2006mi,Yamazaki:2007oc}.
Therefore, one  expects that there is only a  small degeneracy between the PMF parameters and the other cosmological parameters. 
Hence, if one wishes to understand  the qualitative nature of how the PMF parameters are constrained by  observations of  LSS,
one is justified in fixing  the other cosmological parameters at their best fit values from the WMAP analysis. 

\begin{figure*}[h]
\includegraphics[width=1.0\textwidth]{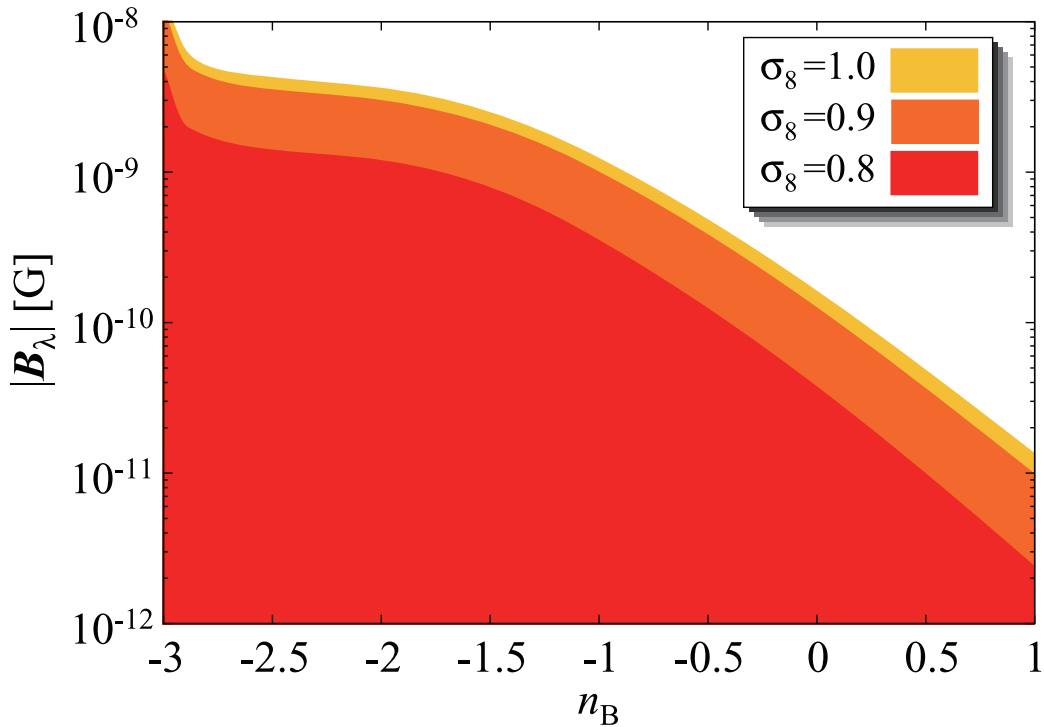}
\caption{\label{sigma8}
Constrained $B_\lambda$ vs. $n_\mathrm{B}$ plane by the $\sigma_8$. 
The deep (red), middle (orange) and pale (yellow) areas are for $\sigma_8 \le 0.8, 0.9$ and 1.0, respectively.
 (For interpretation of the references to colour in this figure legend, the reader is referred to the web version of this article.)
}
\end{figure*}

In Fig. \ref{sigma8} we show the optimum PMF parameters $n_\mathrm{B}$ and $B_\lambda$ for various constant values of $\sigma_8$.
The power spectrum of the PMF $P_{PMF}(k) $ scales  \cite{Yamazaki:2007oc} as: 
\begin{eqnarray}
P_{PMF}(k) \propto k^{2n_\mathrm{B}+3}.
\end{eqnarray}
Therefore, for $n_\mathrm{B} < -1.5$,  the effect of a PMF on the density fluctuations on small scales decreases with smaller values for $n_\mathrm{B}$.
In the case that  $n_\mathrm{B}$ is near $ -3.0$, the matter power spectrum including the PMF effects diminishes  for smaller scales. 
Hence, larger amplitudes of $B_\lambda$ are allowed.
However, for $n_\mathrm{B} > -1.5$  the PMF power spectrum is proportional to  the cut-off scale as $k_\mathrm{C}^{2n_\mathrm{B}+3}$\cite{Yamazaki:2007oc} where $k_\mathrm{C}$  is proportional to $B_\lambda^{-1/(n_\mathrm{B}+5)}$ \cite{Jedamzik:1996wp,Subramanian:1997gi,Banerjee:2004df,Yamazaki:2007oc}.
Substituting these relations into Eqs.(15)-(17) in Ref. \cite{Yamazaki:2007oc}, an expression for the relation between the power spectrum of the PMF and the magnetic strength can be deduced,  
\begin{eqnarray}
P(k)_\mathrm{PMF} \propto B_\lambda^{{[14}/{(n_\mathrm{B}+5})]}~~.
\end{eqnarray}
When $B_\lambda << 1$nG, 
the matter power spectrum including the effect of the PMF for $n_\mathrm{B} > -1.5$ becomes larger for larger $n_\mathrm{B}$.
In this case a large  strength of $B_\lambda$ is not allowed for larger $n_\mathrm{B}$.

If the PMF has no correlation with the primary density fluctuations, 
the effect of a PMF is to increase  the matter power spectrum independently of whether the PMF pressure or tension dominates \cite{Yamazaki:2007oc}.
As noted above,  recent  cosmological observations of LSS 
 imply $\sigma_8 > 1$ can be excluded.
Therefore, we can exclude field strengths of  $B_\lambda\ ^>_\sim 1$ nG if $n_\mathrm{B} > -0.9$ (Fig.~\ref{sigma8}) and $B_\lambda\ ^>_\sim 0.1$ nG if $n_\mathrm{B} > 0.2$. 
When the PMF evolves to the observed magnetic field in clusters of galaxies, we can expect that the PMF amplitude is $B_\lambda\ ^>_\sim 1$ nG. 
In this case from Fig.\ref{sigma8} we can exclude spectral indices of  $n_\mathrm{B} > -0.9$.

\subsection{\label{s3_4}Concordance MCMC analysis}
Constraints  on parameters of the PMF have been deduced by our group \cite{2010PhRvD..81b3008Y} and other groups \cite{2010PhRvD..82h3005K,2011PhRvD..83l3533P}.  These constraints are based upon fits to the CMB and LSS observational data in the context of a flat Lambda CDM cosmology.   This cosmology is characterized by six standard parameters
i.e.~$\{ \Omega_b h^2, 
	\Omega_c h^2, 
	\tau_C, 
	n_s, $
$\log(10^{10}A_s), 
	A_t/A_s$,  
$|B_\lambda|$, 
$n_\mathrm{B} \}$, 
where
$\Omega_b h^2$
and
$\Omega_c h^2$
are the baryon and CDM densities in units of the critical density,
$h$ denotes the present Hubble parameter in units of 100 km s$^{-1}$Mpc$^{-1}$,
$\tau_C$
is the optical depth for Compton scattering,
$n_s$
is the spectral index of the primordial scalar fluctuations,
$A_s$
is the scalar amplitude of the primordial scalar fluctuations and
$A_t$
is the scalar amplitude of the primordial tensor fluctuations.
We define the tensor index for the primordial tensor fluctuations as $n_t =-(A_s/A_t)/8 $.
For these  standard parameters one can adopt the Bayesian priors used in the previous WMAP analysis \cite{2007ApJS..170..377S,2007ApJS..170..288H,2007ApJS..170..335P,Dunkley:2008ie} without a PMF.   To the standard cosmological parameters, one then  adds the two PMF parameters: the  spectral index $n_\mathrm{B}$; and  the average magnetic field strength on a comoving scale  $B_\lambda$.

Using a MCMC method with cosmological observations (e.g. the CMB and/or  LSS) the standard cosmological parameters and the PMF parameters have been constrained \cite{2010PhRvD..81b3008Y}. The results of this analysis  are summarized  in Table \ref{CMBonlyTable}.
However, we note that an  inherent  flaw in  the MCMC approach is  a sparse sampling of points near the boundary.  Consequently, this method tends to assign low probability near a boundary.
This is the reason that  the PMF power law index $n_B$ seems to be constrained even in the limit of $B_\lambda \rightarrow 0$.
In reality, of course, the  spectral index $n_\mathrm{B}$ is not  constrained for  $B_\lambda = 0$.
However, this shortcoming does not negate the result that a finite magnetic field and spectral index give a genuine minimum likelihood in the goodness of fit.
In the best fit including the PMF parameters, the minimum total $\chi^2$ is improved from 2803.4 to 2800.2 corresponding to an improvement of the reduced $\chi^2$ from 1.033 to 1.031. 
This slight  improvement implies that   the existence of the PMF is still only of marginal significance.

\begin{figure*}[h]
\includegraphics[width=0.9\textwidth]{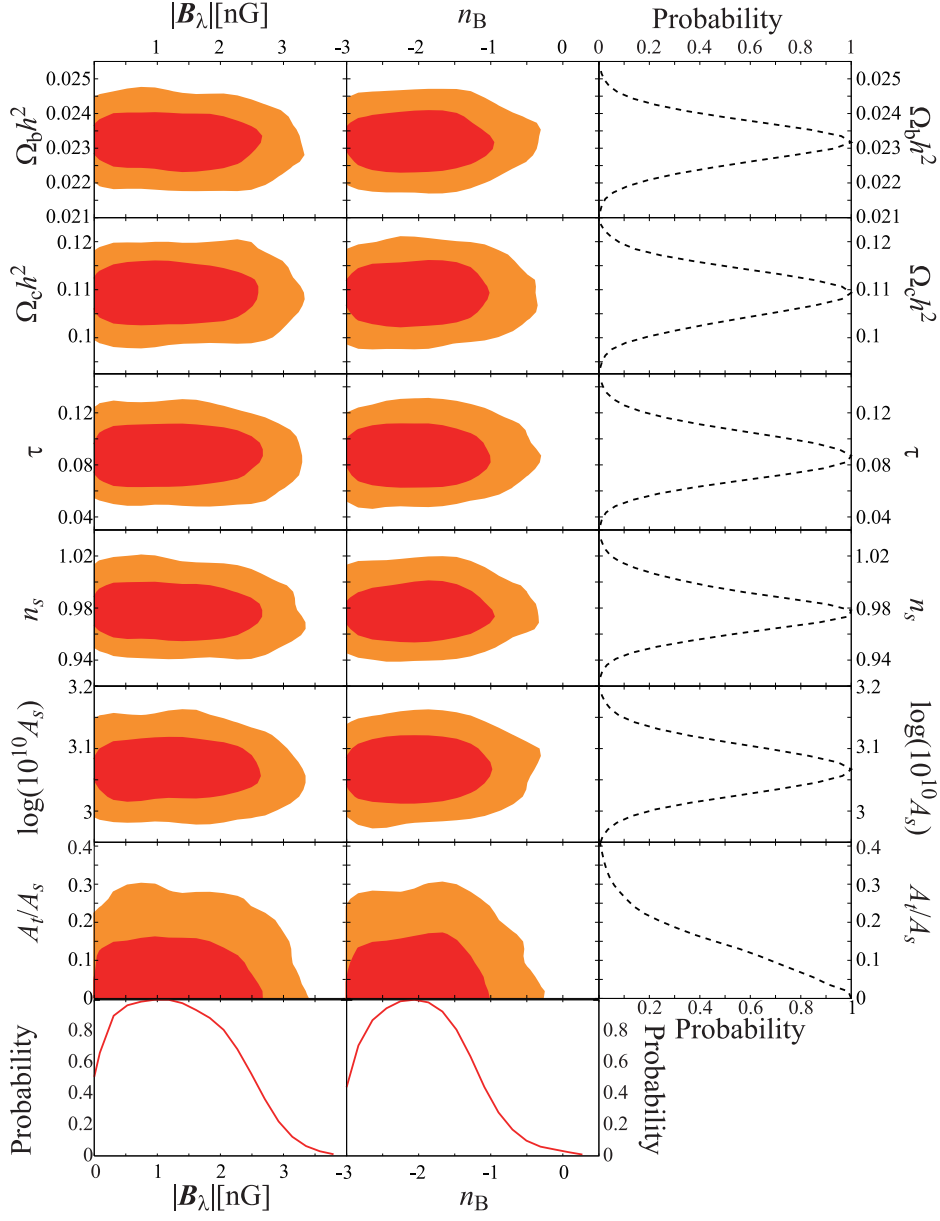}
\caption{\label{c_fsp}
Probability contour plane of PMF parameters ($n_\mathrm{B}$ and $B_\lambda$) vs. the standard cosmological parameters and probability distributions for PMF parameters and the standard cosmological parameters
Deep (red) and pale (orange) contours show the 1 $\sigma$(68\%)
 and 2 $\sigma$(95\%) confidence limits. 
Bold (red) curves on the bottom of the figure are the probability distributions for the PMF parameters.  
Dashed (black) curves on the right of the figure are the probability distributions of each standard cosmological parameter.  
 (For interpretation of the references to colour in this figure legend, the reader is referred to the web version of this article.)
}
\end{figure*}

\begin{figure*}[h]
\includegraphics[width=1.0\textwidth]{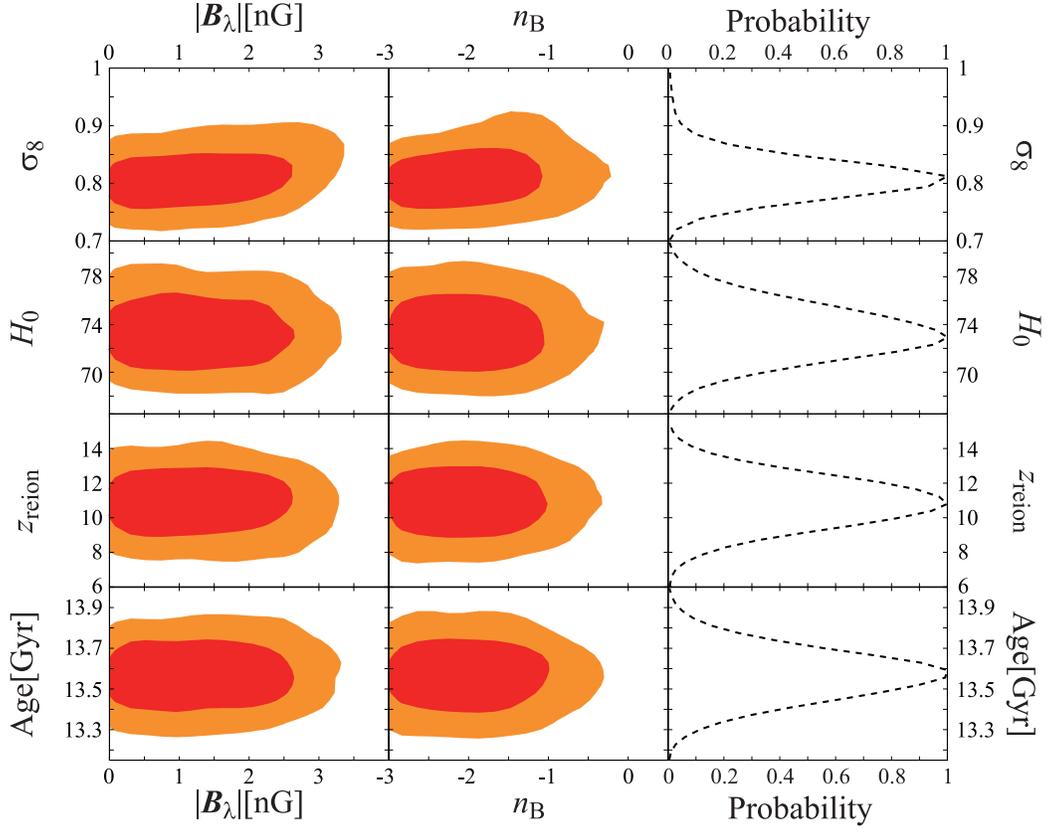}
\caption{\label{c_dp}
Probability contour plane of PMF parameters ($n_\mathrm{B}$ and $B_\lambda$) vs.the derived parameters  $\sigma_8$, $H_0$, $z_\mathrm{reion}$ and Age and   
probability distributions for the derived parameters. 
Deep (red) and pale (orange) contours show the 1 $\sigma$(68\%)
 and 2 $\sigma$(95\%) confidence limits. 
Dashed (black) curves on the right of the figure are the probability distributions for each derived parameter.  
 (For interpretation of the references to colour in this figure legend, the reader is referred to the web version of this article.)
}
\end{figure*}

Figures \ref{c_fsp} and \ref{c_dp} show the 68\% and 95\% C.L.~probability contours in the planes of each standard cosmological parameter versus field strength of the PMF or power law index.  Also shown are  the probability distributions.
The bottom panels of Fig.~\ref{c_fsp} and Fig.~\ref{c_dp} show the probability distributions  for $|B_\lambda|$  and $n_\mathrm{B}$.  
Of particular note for this  MCMC analysis is that a parameter set of  $|B_\lambda| = 0.85 \pm 1.25$ nG and  $n_\mathrm{B}= -2.37^{+0.88}_{-0.73}$ maximizes the likelihood.
These values of the PMF parameters are consistent with no magnetic field, and thus provide  only imply upper limits.  Nevertheless, they suggest that it may be possible  to detect a finite PMF with forthcoming data (particularly for large CMB multipoles).

These figures also confirm that there is  no degeneracy between the standard cosmological parameters and the PMF. We can see this  from Table \ref{CMBonlyTable} in which it is apparent that the standard cosmological parameters do not significantly differ from those deduced directly from the WMAP data without the introduction of a PMF.
As mentioned above, the reason for this is simple.
The standard cosmological parameters are mainly derived from  the observed CMB for low multipoles $\ell < 1000$ (up to the 2nd acoustic peak), while the PMF is most important  for $\ell > 1000$. 
Thus, the PMF effect on the CMB power spectrum is nearly independent of the standard cosmological parameters.

The tensor to scalar ratio $A_t/A_s$ in our analysis is defined as the primary tensor amplitude without the PMF. When we compare our tensor amplitude with the cosmological observations by the MCMC method, we combine the tensor amplitude from the PMF $A_{t\mathrm{[PMF]}}$ and $A_t$.
Therefore the upper limit to the tensor to scalar ratio $A_t/A_s$ in our analysis is less  than the $A_t/A_s$ ratio constrained from the WMAP analysis without a PMF ($A_t/A_s <$ 0.43, 95\% CL). 
Actually, the upper limit of $(A_t+ A_{t\mathrm{[PMF]}})/A_s$ constrained by our analysis including a  PMF  is consistent with the previous constraints.
In addition, the effects of a PMF on the matter power spectrum and the CMB temperature fluctuations and polarization anisotropies dominate for smaller scale e.g. $\ell> 1000$, while the PMF contribution to the CMB for larger scales is  negligible compared to the primary CMB fluctuations of the scalar and tensor modes \cite{Paoletti:2008ck,Finelli:2008xh}. 
Therefore, the tensor to scalar ratio is not affected by the presence of a PMF.

Figure \ref{c_dp} shows the deduced probability distributions and the 1$\sigma$ and $2\sigma$ (68\% and 95\% C.L.)
probability contours for the derived parameters, $\sigma_8$, $H_0$, $z_\mathrm{reion}$, and Age, where $H_0$ is the Hubble parameter  in units  of  km s$^{-1}$ Mpc$^{-1}$, $z_\mathrm{reion}$ is the red shift at which re-ionization occurs, and Age is the presently observed age of the universe in Gyr.
In Fig.~\ref{c_dp}, one can see by the slight bend in the contours that there is a weak degeneracy between the PMF parameters and $\sigma_8$. The reason for this is that the matter power spectrum for smaller scale  determines both $\sigma_8$ and the PMF parameters. 

Table~\ref{CMBonlyTable} summarizes the constrained PMF parameters together with the standard and derived  cosmological parameters.  
The best constraints on the PMF determined yet to date determined in our analyses are \cite{2010PhRvD..81b3008Y}: 
\begin{eqnarray}
|B_\lambda| \mathbf{< 2.10} {\rm~ nG}~( 68\% \mathrm{CL})
 ~,~\mathbf{< 2.98} {\rm~ nG}~( 95\% \mathrm{CL})
\end{eqnarray}
on a present scale of 1 Mpc,  and  
\begin{eqnarray}
n_\mathrm{B} \mathbf{< -1.19}~( 68\% \mathrm{CL})~,~~\mathbf{< -0.25}~( 95\% \mathrm{CL}).
\end{eqnarray}
This result differs only slightly from the constraints obtained in  \cite{2010arXiv1006.4242S,2011PhRvD..83l3533P} based upon different observations.
 
\begin{figure*}[h]
\includegraphics[width=1.0\textwidth]{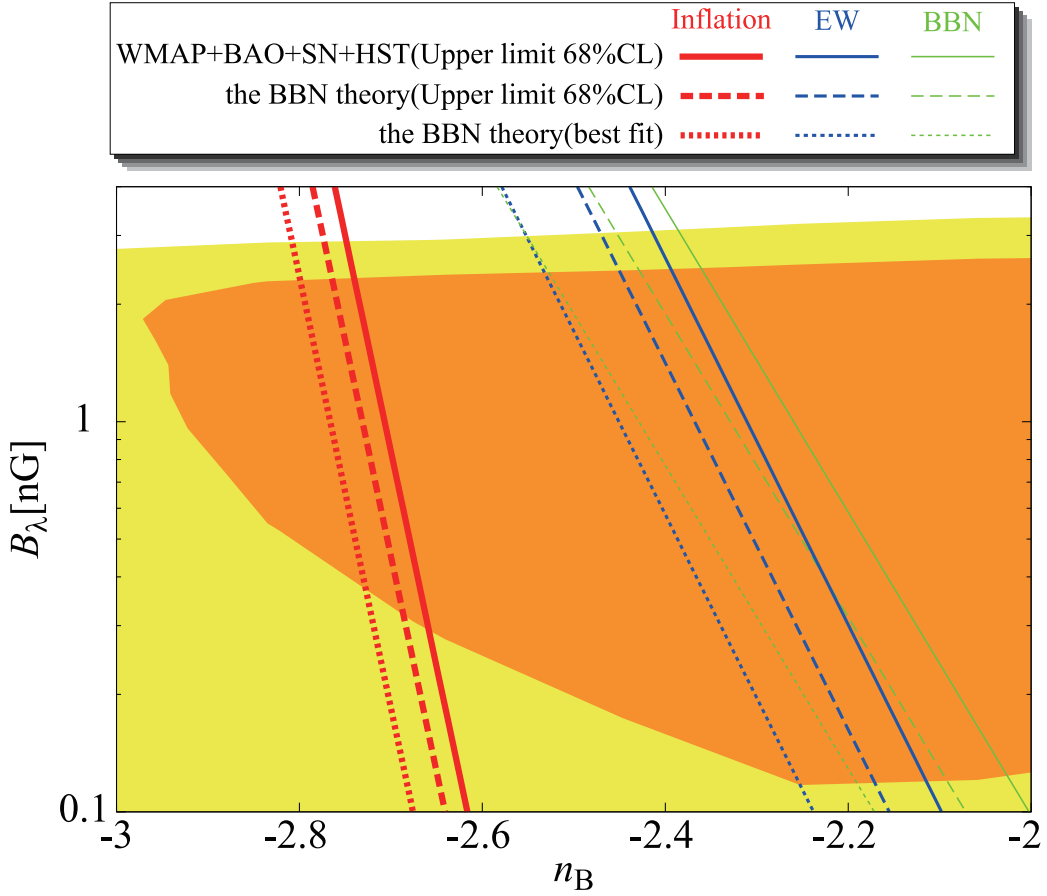}
\caption{\label{c_log_bn}
Allowed and excluded regions in the $1\sigma$ (68\%) C.L. and $2\sigma$ (95.4\%) C.L. in the plane of $|\mathbf{B}_\lambda|$ vs. $n_B$ obtained by an MCMC method applied to  the CMB (WMAP 5yr \cite{Hinshaw:2008kr},
ACBAR\cite{Kuo:2006ya},
CBI\cite{Sievers:2005gj}, 
Boomerang \cite{Jones:2005yb}) and the LSS (2dFDR \cite{Cole:2005sx}) observational data. 
The thin (green), middle (blue), and bold (red) lines show upper limits on the produced PMF during the epoch of BBN, the electroweak transition, and the inflation epoch, respectively.
 (For interpretation of the references to colour in this figure legend, the reader is referred to the web version of this article.)
}
\end{figure*}

Adopting these PMF constraints and the BBN GWB constraint  discussed  above \cite{Caprini:2001nb},  a re-examination  of the  three main PMF generation scenarios
\cite{Vachaspati:1991nm,Kibble:1995aa,Ahonen:1997wh,Joyce:1997uy}
  is motivated by the present results.
Figure~\ref{c_log_bn} summarizes the contours of $|B_\lambda| $ and $n_\mathrm{B}$ at the 1 and 2 $\sigma$ C.L. along with the associated GWB constraints for the three generation epochs.

The region bounded by the yellow area corresponds to  the  2$\sigma$ C.L.  on  PMF parameters.
The upper limit of the produced PMF from big-bang nucleosynthesis, the electroweak transition, and the inflation epoch are shown by the green, blue, red lines, respectively.
However, as mentioned above, there is not yet a  lower limit for the  PMF parameters, $B_\lambda$ and $n_\mathrm{B}$ at the 2$\sigma$ C.L.
On the other hand, considering the region bounded by the $\sigma$ C. L. red orange area, one can find both upper and lower limits to   the PMF strength.
Figure. \ref{c_log_bn} shows that the allowed or excluded regions based upon these multiple constraints depends upon when the PMF was generated as follows:\footnote{The 1 $\sigma$ and 2$\sigma$ regions on the two parameter plane like Fig.\ref{c_log_bn} do not directly indicate constrained values of the parameters at 1$\sigma$ and 2$\sigma$ as given in Table~\ref{CMBonlyTable}}

At the  2$\sigma$ C.L. there are only upper limits to  $|B_\lambda|$ and no limits on $n_B$.  These are: 

I. Inflation:
\begin{eqnarray}
|B_\lambda|~\lesssim~ 2.90~\mathrm{nG} \nonumber\\
\end{eqnarray}

II. Electroweak transition:
\begin{eqnarray}
|B_\lambda|~\lesssim~ 3.08~\mathrm{nG} \nonumber\\
\end{eqnarray}

III. BBN:
\begin{eqnarray}
|B_\lambda|~\lesssim~ 3.10~\mathrm{nG} \nonumber\\
\end{eqnarray}

At the  1$\sigma$ C.L. there are both upper lower limits to $|B_\lambda|$ and  $n_B$.  These are: 

I. Inflation:
\begin{eqnarray}
0.292\ \mathrm{nG}\ \lesssim\ |B_\lambda|\ \lesssim\ 2.33~\mathrm{nG} \nonumber\\
-2.97\ \lesssim\ n_B\ \lesssim\ -2.66 \nonumber
\end{eqnarray}

II. Electroweak transition:
\begin{eqnarray}
0.117\ \mathrm{nG}\ \lesssim\ |B_\lambda|\ \lesssim\ 2.46~\mathrm{nG} \nonumber\\
-2.97\ \lesssim\ n_B\ \lesssim\ -2.11 \nonumber
\end{eqnarray}

III. BBN:
\begin{eqnarray}
0.117\ \mathrm{nG}\ \lesssim\ |B_\lambda|\ \lesssim\ 2.48~\mathrm{nG} \nonumber\\
-2.97\ \lesssim\ n_B\ \lesssim\ -2.03 \nonumber
\end{eqnarray}

If the PMF were  generated at an even earlier epoch, 
it is clear that  upper limits on both  $|B_\lambda|$ and $n_B$ would be  more stringent.
These  limits are the strongest constraints on the PMF that have yet been determined. 
However, one should keep in mind that the evolution and/or the generation of the magnetic field on cosmological scales during the formation of LSS is not well understood. 
Therefore, if there are other effective physical scenarios for the generation and/or evolution of the PMF after the epoch of the photon last scattering,
 our lower limits to the PMF parameters will change.
To accurately constrain the PMF parameters, 
one  should study the PMF not only before but also after the epoch of  photon last scattering.

\begin{table}
\begin{tabular}{ccc} 
\multicolumn{3}{c}{\it Cosmological Parameters}\\
\hline
\multicolumn{1}{c}{Parameter} &
\multicolumn{1}{c}{
mean}&
\multicolumn{1}{c}{best fit}\\
\hline
$\Omega_b h^2$ &
$0.02320 \pm 0.00059$ &
$0.02295$ \\
$\Omega_c h^2$ &
$0.1094\pm0.0046$ &
$0.1093$ \\
$\tau_C$ &
$0.087\pm0.017$ &
$0.082$ \\
$n_s$ &
$0.977\pm0.016$ &
$0.970$ \\
$\ln(10^{10}A_s)$ &
$3.07\pm0.036$ &
$ 3.06$\\
$A_t/A_s$ &
$< 0.170 ( 68\% \mathrm{CL}), < 0.271 ( 95\% \mathrm{CL})$ &
$0.0088$\\
$|B_\lambda|\mathrm{(nG)}$ &
$\mathbf{< 2.10} ( 68\% \mathrm{CL}), \mathbf{< 2.98} ( 95\% \mathrm{CL})$ &
$\mathbf{0.85}$\\
$n_\mathrm{B}$ &
$\mathbf{< -1.19} ( 68\% \mathrm{CL}), \mathbf{< -0.25} ( 95\% \mathrm{CL})$ &
$\mathbf{-2.37}$\\
\hline
$\sigma_8$ &
$0.812^{+0.028}_{-0.033}$ &
$ 0.794$\\
$H_0$ &
$73.3\pm2.2$ &
$ 72.8$\\
$z_\mathrm{reion}$ &
$10.9\pm 1.4$ &
$ 10.5$\\
$\mathrm{Age(Gyr)}$ &
$13.57\pm 0.12$ &
$ 13.62$\\
\hline\hline
\end{tabular}
\caption{PMF parameters and $\Lambda$CDM model parameters and 68\%
confidence intervals ($A_t/A_s$  is a 95\% CL) from a fit to the seven year  WMAP \cite{Hinshaw:2008kr} + ACBAR \cite{Kuo:2006ya} + CBI \cite{Sievers:2005gj} +Boomerang \cite{Jones:2005yb} +  2dFDR \cite{Cole:2005sx} data. }
\label{CMBonlyTable}
\end{table}
\section{\label{s4}Discussion and Future Challenges}
The generation and/or the  evolution of the PMF along with its  physical behavior in the early universe have been studied by  many researchers [see \cite{2011PhR...505....1K} for a recent review].
In this section we review several relevant  topics and their implications for different PMF strengths.
We  also discuss  future prospects  for  studying the role of  the PMF in  cosmology, astrophysics and astronomy.

\subsection{\label{s4_1} The BB mode from the PMF}
\begin{figure*}[h]
\includegraphics[width=1.0\textwidth]{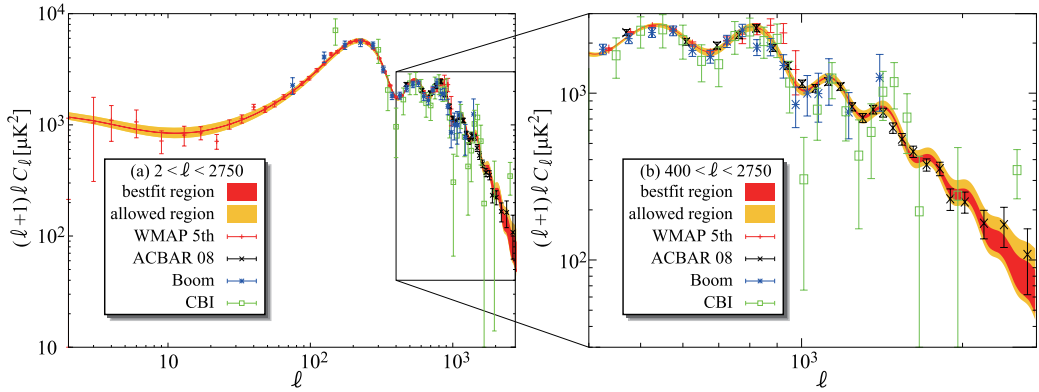}
\caption{\label{TTbest}
Comparison of the observed CMB with the computed total TT power spectra from the best-fit and allowed parameters. Various ranges of each plats are (a) TT($2 < \ell < 2750$) and (b) TT($400 < \ell < 2750$). Deep colour (red) regions indicate the best-fit parameter set and allowed regions [pale (orange) colour] are from the constrained parameter set of Table.1. Dots with error bars show the WMAP 5yr, ACBAR 08, Boomerang and CBI data as the legend box in the each panel. The upper curves of allowed regions in each panel include the SZ effect at the K(22.8GHz) band, and the lower curves do not include the SZ effect. (For interpretation of the references to colour in this figure legend, the reader is referred to the web version of this article.) (For interpretation of the references to colour in this figure legend, the reader is referred to the web version of this article.)
}
\end{figure*}

\begin{figure*}[h]
\includegraphics[width=1.0\textwidth]{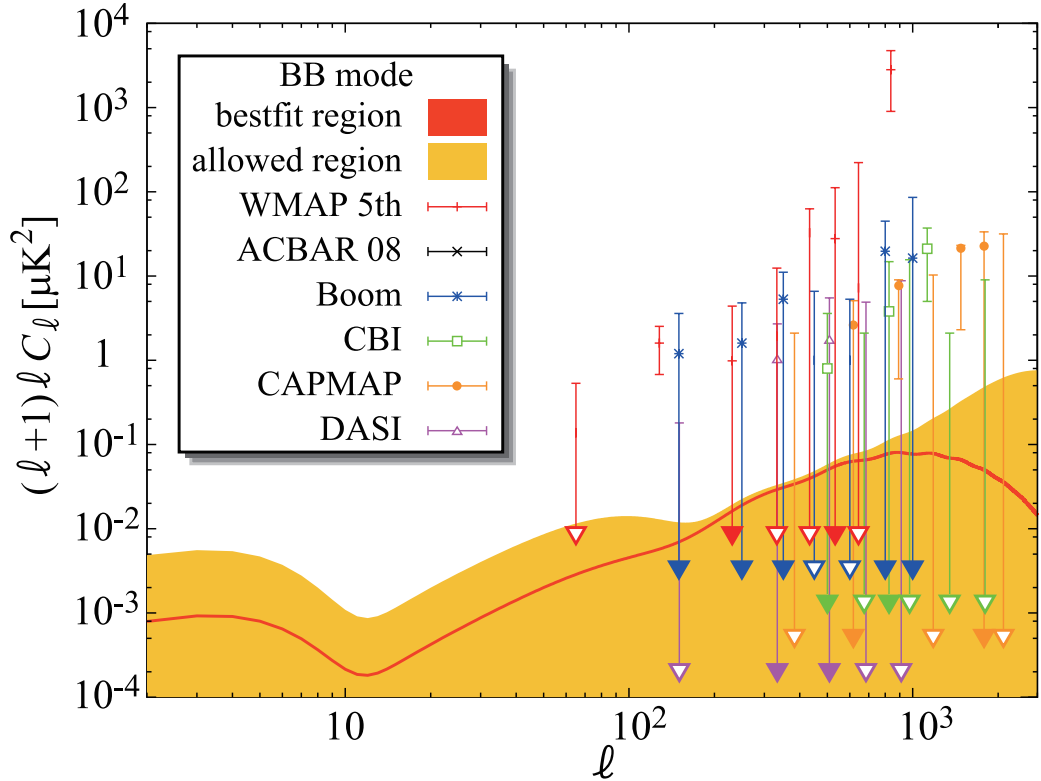}
\caption{\label{BBbest} 
Comparison of the observed CMB with the computed total BB power spectra from the best-fit and allowed parameters.
Deep colour (red) regions indicate the best-fit parameter set
and allowed regions [pale (orange) colour] are from the constrained parameter set of Table.1. 
Dots with error bars show the WMAP 5yr, ACBAR 08, Boomerang, CBI, CAPMAP, and DASI data as the legend box in the figure.
Downward arrows for the error bars indicate that the data points are upper limits.
 (For interpretation of the references to colour in this figure legend, the reader is referred to the web version of this article.)
}
\end{figure*}

As mentioned above, a PMF affects not only the temperature fluctuations, but also the polarization anisotropies of the CMB. 
Although  fits have been based upon  all available temperature and polarization data of the CMB, it turns out that the TT and BB modes are the most important for constraining the parameters of the PMF.
  Figs.~\ref{TTbest} (the TT-mode) and \ref{BBbest} (the BB-mode) show a comparison of the observations of the CMB with the computed best-fit total power spectrum.
 Figure \ref{TTbest} shows the best fit and allowed regions including both the Sunyaev Zel'dovich (SZ) effect (scattering from re-ionized electrons) at the K(22.8GHz) band (upper curves) and without the SZ effect (lower curves).  Including the SZ effect only slightly diminishes the best fit magnitude of the PMF.

 Figure \ref{TTbest}  exhibits the best-fit and allowed regions (or curve) both including 
the SZ effect at 22.8GHz (upper curves) and without the SZ effect (lower curves). 
Since the temperature fluctuations of the CMB from recent observations have become more accurate than the data shown in Fig.~\ref{TTbest}, these observations give a strong prior range for constraining parameters of the PMF.
However, as discussed above, 
we cannot determine the parameters of the PMF by the TT mode alone because there is a strong degeneracy between  the field strength and the power law index of the PMF.
On the other hand, the effects of a PMF on the BB-mode of the CMB has a feature for $\ell > 200$ as shown in Fig. \ref{BBbest}. This  is due to the vector source from the PMF\cite{Yamazaki:2006bq,Yamazaki:2007oc}.
Thus, if accurate  observations of the BB mode could  be obtained in the future, one expects that the parameters of the PMF could be constrained without the uncertainty due degeneracy between  the field strength and the power law index.

\subsection{\label{s4_2} PMF and GWB}
Several groups \cite{Caprini:2001nb,2010PhRvD..81b3002W}  have studied the GWB induced  by a PMF.
From such studies, one expects that there is a degeneracy between the GWB from  inflation and that generated by a  PMF.
Indeed, if the PMF is generated during the  inflation epoch this degeneracy would become stronger.  This is particularly true since  the power spectrum of an inflation-generated  PMF is expected to be  nearly scale-invariant. 
Therefore, even if the GWB is  observed in the future, it will still be difficult to constrain the parameters of the PMF from these observations alone(See Fig.12).

  There is, however, a difference.  The GWB from inflation  mainly affects the tensor BB mode of the CMB, 
while a PMF affects both the vector and tensor polarizations of the BB mode of the CMB \cite{Yamazaki:2007oc,2010PhRvD..81b3008Y}. Thus, the BB mode exhibits different  features in the GWB from a PMF or an  inflation origin. 
Since  both a  PMF and  inflation dominate the tensor source  on the same scales of the BB mode, there is  a strong degeneracy between the BB mode of tensor type polarizations from the PMF and  inflation. However, the PMF also has a vector type polarization as a source of the BB mode of the CMB and this mode is most important   on smaller angular scale e.g. $\ell~>~200$, while both tensor sources  of the PMF and inflation dominate the BB mode on larger angular scales e.g. $\ell~>~200$. 
Therefore, if one could simultaneously obtain observations of the GWB and the tensor and vector BB modes of the  CMB,  this degeneracy may be resolved.
\subsection{\label{s4_3}Constraint on the  neutrino mass in the presence of a PMF}
The presence of a PMF can alter the particle constraints deduced from the CMB.  If the velocities of finite-mass neutrinos are large enough, the growth of density fluctuations is impeded on the free-streaming-scale of the neutrinos \cite{1980PhRvL..45.1980B}. Hence, the matter density field in the early universe can be affected by a finite neutrino mass \cite{Lesgourgues:2006nd}.  The inferred  cosmological constraints on the mass of the neutrino are of considerable interest. 
The sum of the neutrino masses is constrained to be $\sum_\nu m_\nu \le 0.1-1$ eV by a combination of cosmological observations \cite{DeBernardis:2008qq,Ichiki:2008rh,Ichiki:2008ye,2011ApJS..192...18K} and the results of  tritium beta-decay endpoint experiments \cite{2008RPPh...71h6201O}. On the other hand, as explained Sec.~\ref{s3_3}, a PMF affects the matter density fields in the early universe and the parameters of the PMF correlate with $\sigma_8$.
Therefore, there is a possible  degeneracy between the parameters of the PMF and the mass of the neutrino.  This degeneracy relaxes the cosmological constraint on the sum of the neutrino masses.

In Fig.~\ref{Neu} we show the constraints on the sum of the neutrino masses $\sum_\nu  m_\nu$ and $B_\lambda$ for various fixed values of $n_\mathrm{B}$ and ranges of $\sigma_8$. 
The expected field strength of the PMF from  cosmological observations is $B_\lambda<$2.0nG$(1 \sigma)$ and $<$3.0nG$(2 \sigma)$, while  $\sigma_8$ is constrained  to be in the range 0.75$< \sigma_8 < $0.85.
For this range of $\sigma_8$, if we do not consider the effect of a PMF on the matter density field, the constrained sum of the neutrino masses is less than 
$0.11$ eV.
On the other hand when a PMF is introduced,  the upper limit on the sum of the neutrino masses from Fig.~\ref{Neu}, increases to  $\sum_{N_\nu} m_\nu < 0.24$ eV for  $n_\mathrm{B}=-1.5$ and $< 0.6$eV  for  $n_\mathrm{B}=-2.5$  and $N_\nu = 3$.  
Since the effect of a PMF cancels the effect of neutrinos on the density fluctuations, 
the constrained mass of the neutrinos when a  PMF is present is larger than that deduced without considering a PMF.

However, $\sigma_8$ depends upon the cosmological model employed and 
it has some  degeneracy with 
$\Omega_m = \Omega_b + \Omega_\mathrm{CDM}$, 
$n_\mathrm{S}$, and 
$A_\mathrm{S}$, even if these primary parameters are well constrained by the CMB data, e.g. WMAP\cite{2011ApJS..192...18K}. 
Furthermore, the neutrino mass has a  degeneracy with $\Omega_m$ \cite{Elgaroy:2002bi}. 
Thus, one should ultimately consider the simultaneous degeneracy between the standard cosmological parameters,  the mass of neutrinos and the PMF.
With high precision  observations on small angular scales, it may be possible in the near future to obtain not only  upper but also  lower limits to the mass of the neutrinos from cosmology and astrophysics in the presence of the PMF.
\subsection{\label{s4_4}Non-Gaussianity from a PMF}
If we assume that the inflation mode is simple, then the primordial fluctuations will be  Gaussian. 
A measurement  of non-Gaussianity in the primordial fluctuations could thus  provide   important  evidence for new physics beyond that of simple inflation\cite{2003ApJS..148..119K}. 
Recently, however, several authors have shown that a PMF can cause  non-Gaussianity\cite{2011PhRvD..83l3003S,2010PhRvD..82l3006T,2009JCAP...06..021C,2009PhRvL.103h1303S,2005PhRvD..72f3002B}.

We can constrain the non-Gaussianity by auto- and cross correlated bispectra from the intensity fluctuations of the CMB. On the other hand, the bispectra from the CMB are also affected by a PMF. Therefore, one expects that the effects of a PMF and inflation  on the non-Gaussianity of the CMB will have a degeneracy.
Hence, even if  one could obtain accurate observations of  the non-Gaussianity of the CMB, one would not be able to analyze the inflation model without  information on the effects from the PMF. To solve this degeneracy and truly understand the inflation model, one should consider the constraints on  PMF effects by various observations, e.g. BBN, the GWB,  LSS, and the CMB.
\subsection{\label{s4_5}21 cm line and a PMF}
The 21 cm line is emitted by a transition of the neutral hydrogen atom
between the two different energy levels of its 1s ground state. Since 
hydrogen is the most abundant element in the Universe (75
\% of normal matter by mass and 90 \% by number), it is
 important to survey the 21 cm line for researching the "Dark
Age"\footnote{Considering the cosmological redshift, the
21 line is observed in a frequency range from about 200MHz to 9 MHz.}. If
observations of the 21 cm are successful, one should be able to obtain a
precise  matter power spectrum after recombination and the re-ionization
of the Universe\cite{2006MNRAS.372.1060T}.

As mentioned above, a PMF affects  structure formation in the early
Universe. Furthermore, a PMF of strength $\sim 1$ nG  reheats the intergalactic medium (IGM) including hydrogen up to several thousand K through  ambipolar
diffusion\cite{Sethi:2008eq}. Therefore, the 21 cm line is indirectly
affected by a PMF. The Square Kilometer Array (SKA) \footnote{the detected
frequency range of the SKA project will be from 70 MHz to 10GHz.} is planning a
most promising future project for observing the cosmological 21 cm line. One
should consider the effects of a PMF on the 21 cm line to obtain an accurate
physical interpretation from the results of observations like that of  the SKA.

\begin{figure*}[h]
\includegraphics[width=1.0\textwidth]{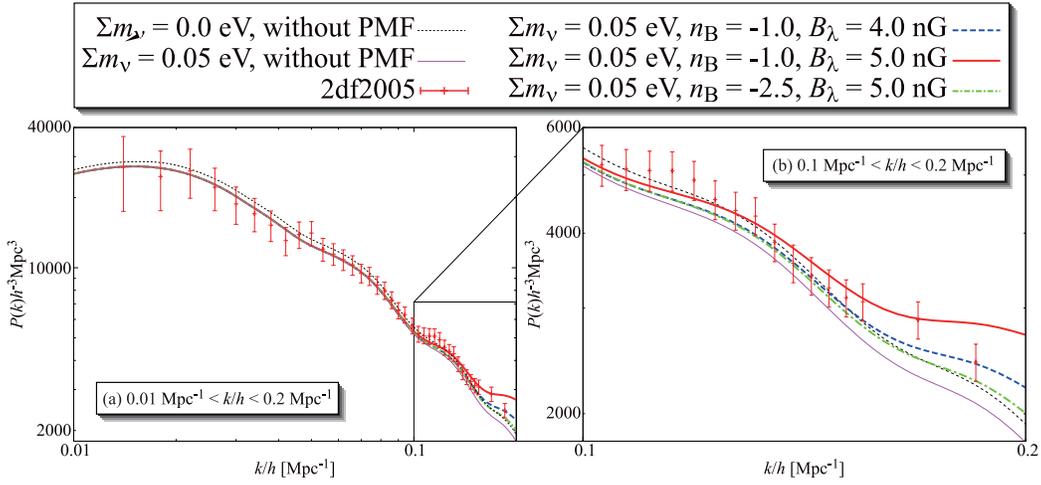}
\caption{\label{Neu_map}
Matter power spectra from finite mass neutrinos and a PMF.
The range of the left and right panel is $0.01 \mathrm{Mpc}^{-1} k/h < 0.2 \mathrm{Mpc}^{-1}$ and $0.1 \mathrm{Mpc}^{-1} k/h < 0.2 \mathrm{Mpc}^{-1}$ 
Dots with error bars and curves are as indicated in the legend box in the figure. 
To better illustrate effects of finite-mass neutrinos and a PMF for the matter power spectra, we have used a larger field strength of a PMF $B_\lambda$.
This figure shows that a PMF affects strongly the matter power spectra on $k/h > 0.1 \mathrm{Mpc}^{-1}$.  The magnitude of PMF effect depends on both $B_\lambda$ and $n_\mathrm{B}$. On one hand, the total amplitudes of the matter power spectra are decreased by the neutrino mass. Furthermore, the effects from the mass of neutrinos on larger $k$ (smaller scales) are greater than on smaller $k$ (larger scales).
}
\end{figure*}

\begin{figure*}[h]
\includegraphics[width=1.0\textwidth]{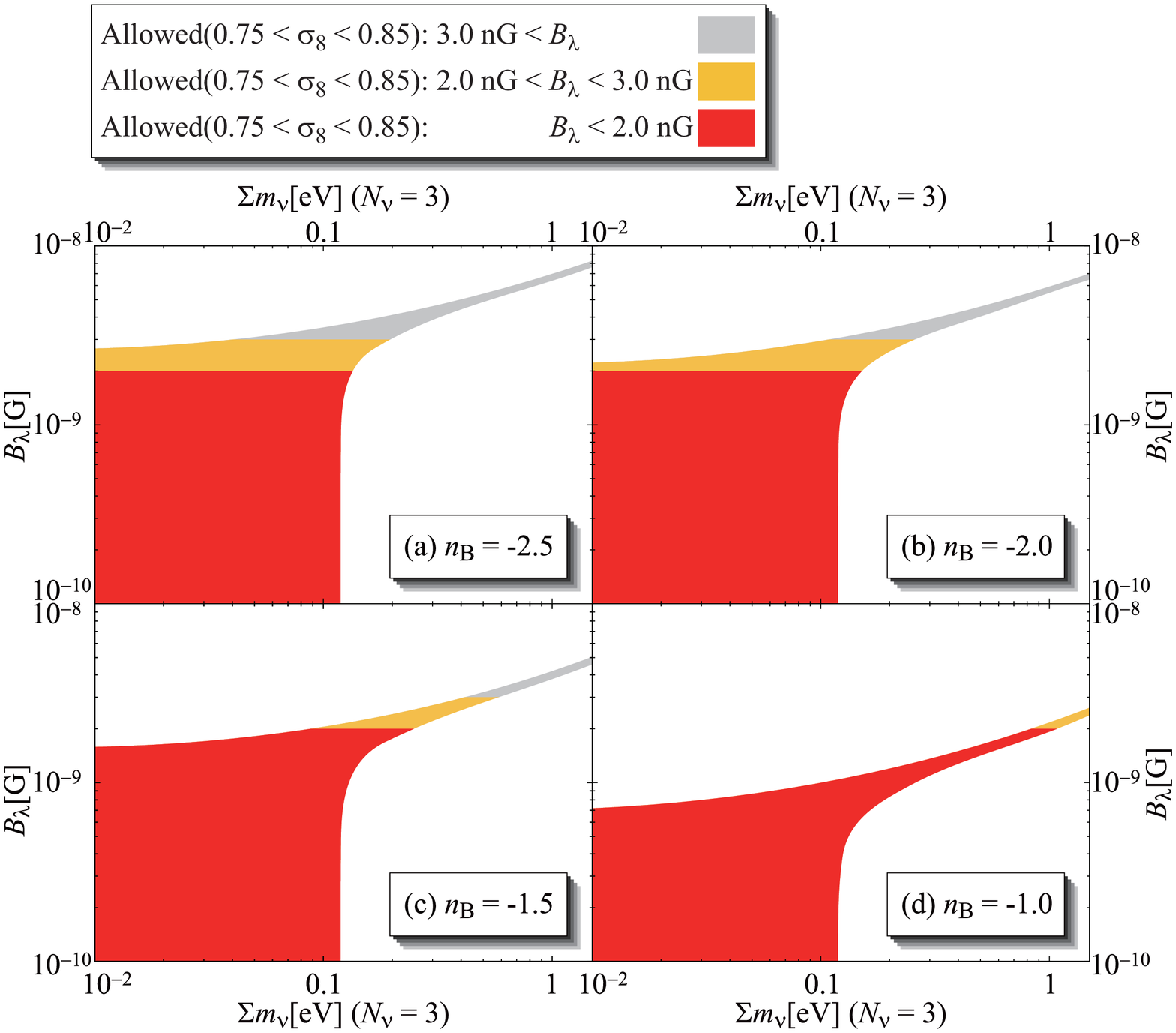}
\caption{\label{Neu}
Allowed and excluded regions for the $\sigma_8$ on the $B_\lambda$ vs. $\sum_{N_\nu = 3} m_\nu $ plane.
Filled regions indicate ranges of $\sigma_8$ as $0.75 < \sigma_8 < 0.85$ and 
pale colour (gray), middle colour (yellow) and deep colour (red) regions show $B_\lambda > 3.0$ nG, $B_\lambda < 3.0$ nG and $B_\lambda < 2.0$nG, respectively.
 (For interpretation of the references to colour in this figure legend, the reader is referred to the web version of this article.)
}
\end{figure*}
\newpage
{\it Acknowledgments} 

This work has been supported in part by Grants-in-Aid for Scientific
Research (20105004, 20244035, 20169444, 21740177 and 22012004) of the Ministry of Education, Culture, Sports,
Science and Technology of Japan.  
This work is also supported by the JSPS Core-to-Core Program, International
Research Network for Exotic Femto Systems (EFES) and Heiwa Nakajima Foundation.
Work at UND supported in part by the US Department of Energy under research grant DE-FG02-95-ER40934.






\bibliographystyle{elsarticle-num}
\bibliography{list}

\begin{thebibliography}{10}
\expandafter\ifx\csname url\endcsname\relax
  \def\url#1{\texttt{#1}}\fi
\expandafter\ifx\csname urlprefix\endcsname\relax\def\urlprefix{URL }\fi
\expandafter\ifx\csname href\endcsname\relax
  \def\href#1#2{#2} \def\path#1{#1}\fi

\bibitem{Subramanian:1998fn}
K.~Subramanian, J.~D. Barrow, Microwave background signals from tangled
  magnetic fields, Phys. Rev. Lett. 81 (1998) 3575--3578.

\bibitem{Mack:2001gc}
A.~Mack, T.~Kahniashvili, A.~Kosowsky, Vector and tensor microwave background
  signatures of a primordial stochastic magnetic field, Phys. Rev. D 65 (2002)
  123004.

\bibitem{Subramanian:2002nh}
K.~Subramanian, J.~D. Barrow, Small-scale microwave background anisotropies due
  to tangled primordial magnetic fields, Mon. Not. Roy. Astron. Soc. 335 (2002)
  L57.

\bibitem{Lewis:2004ef}
A.~Lewis, Cmb anisotropies from primordial inhomogeneous magnetic fields, Phys.
  Rev. D 70 (2004) 043011.

\bibitem{Yamazaki:2004vq}
D.~G. Yamazaki, K.~Ichiki, T.~Kajino, Constraining primordial magnetic field
  from cmb anisotropies at higher multipoles, Astrophys. J. 625 (2005) L1--L4.

\bibitem{Kahniashvili:2005xe}
T.~Kahniashvili, B.~Ratra, Effects of cosmological magnetic helicity on the
  cosmic microwave background, Phys. Rev. D 71 (2005) 103006.

\bibitem{Challinor:2005ye}
A.~Challinor, Cosmic microwave background polarization analysis, Lect. Notes
  Phys. 653 (2004) 71--104.

\bibitem{Dolgov:2005ti}
A.~D. Dolgov, Cosmological magnetic fields and cmbr polarization\href
  {http://arxiv.org/abs/astro-ph/0503447} {\path{arXiv:astro-ph/0503447}}.

\bibitem{Gopal:2005sg}
R.~Gopal, S.~K. Sethi, Tangled magnetic fields and cmbr signal from
  reionization epoch, Phys. Rev. D 72 (2005) 103003.

\bibitem{Yamazaki:2005yd}
D.~G. Yamazaki, K.~Ichiki, T.~Kajino, Primordial magnetic field at the photon
  last scattering surface, Nuclear Physics A 758 (2005) 791--794.

\bibitem{Kahniashvili:2006hy}
T.~Kahniashvili, B.~Ratra, {CMB anisotropies due to cosmological magnetosonic
  waves}, Phys. Rev. D75 (2007) 023002.

\bibitem{Yamazaki:2006bq}
D.~G. Yamazaki, K.~Ichiki, T.~Kajino, G.~J. Mathews, Constraints on the
  evolution of the primordial magnetic field from the small-scale cosmic
  microwave background angular anisotropy, Astrophys. J. 646 (2006) 719--729.

\bibitem{Yamazaki:2006ah}
D.~G. Yamazaki, K.~Ichiki, T.~Kajino, G.~J. Mathews, Primordial magnetic field
  constrained from cmb anisotropies,and its generation and evolution before,
  during and after the bbn, PoS(NIC-IX). (2006) 194.

\bibitem{Giovannini:2006kc}
M.~Giovannini, Tight coupling expansion and fully inhomogeneous magnetic
  fields, Phys. Rev. D 74 (2006) 063002.

\bibitem{Yamazaki:2007oc}
D.~G. {Yamazaki}, K.~{Ichiki}, T.~{Kajino}, G.~J. {Mathews}, {Effects of a
  primordial magnetic field on low and high multipoles of the cosmic microwave
  background}, Phys. Rev. D 77 (2008) 043005.

\bibitem{Paoletti:2008ck}
D.~Paoletti, F.~Finelli, F.~Paci, {The full contribution of a stochastic
  background of magnetic fields to CMB anisotropies}, Mon. Not. Roy. Astron.
  Soc. 396 (2009) 523--534.

\bibitem{Yamazaki:2008bb}
D.~G. Yamazaki, K.~Ichiki, T.~Kajino, G.~J. Mathews, Constraints on the
  primordial magnetic field from $\sigma_8$, Phys. Rev. D 78 (2008) 123001.

\bibitem{2008nuco.confE.239Y}
D.~G. {Yamazaki}, K.~{Ichiki}, T.~{Kajino}, G.~J. {Mathews}, {A Strong
  Constraint on the Neutrino Mass from the Formation of Large Scale Structure
  in the Presence of the Primordial Magnetic Field}, PoS(NIC-X). (2008) 239.

\bibitem{Sethi:2008eq}
S.~K. Sethi, B.~B. Nath, K.~Subramanian, {Primordial magnetic fields and
  formation of molecular hydrogen}, Mon. Not. Roy. Astron. Soc. 387 (2008)
  1589--1596.

\bibitem{Kojima:2008rf}
K.~Kojima, K.~Ichiki, D.~G. Yamazaki, T.~Kajino, G.~J. Mathews, {Neutrino mass
  effects on vector and tensor CMB anisotropies in the presence of a primordial
  magnetic field}, Phys. Rev. D78 (2008) 045010.

\bibitem{2008PhRvD..78f3012K}
T.~{Kahniashvili}, G.~{Lavrelashvili}, B.~{Ratra}, {CMB temperature anisotropy
  from broken spatial isotropy due to a homogeneous cosmological magnetic
  field}, Phys. Rev. D78 (2008) 063012.

\bibitem{Giovannini:2008aa}
M.~Giovannini, K.~E. Kunze, {Faraday rotation, stochastic magnetic fields and
  CMB maps}, Phys. Rev. D78 (2008) 023010.
\newblock \href {http://arxiv.org/abs/0804.3380} {\path{arXiv:0804.3380}},
  \href {http://dx.doi.org/10.1103/PhysRevD.78.023010}
  {\path{doi:10.1103/PhysRevD.78.023010}}.

\bibitem{2010PhRvD..81b3008Y}
D.~G. {Yamazaki}, K.~{Ichiki}, T.~{Kajino}, G.~J. {Mathews}, {New constraints
  on the primordial magnetic field}, Phys. Rev. D 81 (2010) 023008.

\bibitem{2010PhRvD..81j3519Y}
D.~G. {Yamazaki}, K.~{Ichiki}, T.~{Kajino}, G.~J. {Mathews}, {Constraints on
  the neutrino mass and the primordial magnetic field from the matter density
  fluctuation parameter ${\sigma}_{8}$}, Phys. Rev. D 81 (2010) 103519.

\bibitem{2010AdAst2010E..80Y}
D.~G. {Yamazaki}, K.~{Ichiki}, T.~{Kajino}, G.~J. {Mathews}, {Primordial
  Magnetic Field Effects on the CMB and Large-Scale Structure}, Advances in
  Astronomy 2010 (2010) 586590.
\newblock \href {http://dx.doi.org/10.1155/2010/586590}
  {\path{doi:10.1155/2010/586590}}.

\bibitem{Sethi:2003vp}
S.~K. Sethi, {Large Scale Magnetic Fields: Galaxy Two-Point correlation
  function}, Mon. Not. Roy. Astron. Soc. 342 (2003) 962.

\bibitem{Sethi:2004pe}
S.~K. Sethi, K.~Subramanian, {Primordial Magnetic Fields in the
  Post-recombination Era and Early Reionization}, Mon. Not. Roy. Astron. Soc.
  356 (2005) 778--788.

\bibitem{Yamazaki:2006mi}
D.~G. Yamazaki, K.~Ichiki, K.~I. Umezu, H.~Hanayama, Effect of primordial
  magnetic field on seeds for large scale structure, Phys. Rev. D 74 (2006)
  123518.

\bibitem{2011PhR...505....1K}
A.~{Kandus}, K.~E. {Kunze}, C.~G. {Tsagas}, {Primordial magnetogenesis},
  Physics Reports, 505 (2011) 1--58.

\bibitem{Turner:1987bw}
M.~S. Turner, L.~M. Widrow, Inflation produced, large scale magnetic fields,
  Phys. Rev. D 37 (1988) 2743.

\bibitem{Ratra:1991bn}
B.~Ratra, Cosmological 'seed' magnetic field from inflation, Astrophys. J. 391
  (1992) L1--L4.

\bibitem{Bamba:2004cu}
K.~Bamba, J.~Yokoyama, Large-scale magnetic fields from dilaton inflation in
  noncommutative spacetime, Phys. Rev. D 70 (2004) 083508.

\bibitem{Vachaspati:1991nm}
T.~Vachaspati, Magnetic fields from cosmological phase transitions, Phys. Lett.
  B265 (1991) 258--261.

\bibitem{Kibble:1995aa}
T.~W.~B. Kibble, A.~Vilenkin, {Phase equilibration in bubble collisions}, Phys.
  Rev. D52 (1995) 679--688.

\bibitem{Ahonen:1997wh}
J.~Ahonen, K.~Enqvist, {Magnetic field generation in first order phase
  transition bubble collisions}, Phys. Rev. D57 (1998) 664--673.

\bibitem{Joyce:1997uy}
M.~Joyce, M.~E. Shaposhnikov, {Primordial magnetic fields, right electrons, and
  the Abelian anomaly}, Phys. Rev. Lett. 79 (1997) 1193--1196.

\bibitem{Takahashi:2005nd}
K.~Takahashi, K.~Ichiki, H.~Ohno, H.~Hanayama, Magnetic field generation from
  cosmological perturbations, Phys. Rev. Lett. 95 (2005) 121301.

\bibitem{Hanayama:2005hd}
H.~Hanayama, et~al., Biermann mechanism in primordial supernova remnant and
  seed magnetic fields, Astrophys. J. 633 (2005) 941.

\bibitem{ichiki:2006sc}
K.~Ichiki, K.~Takahashi, H.~Ohno, H.~Hanayama, N.~Sugiyama, Cosmological
  magnetic field: a fossil of density perturbations in the early universe,
  Science 311 (2006) 827.

\bibitem{2004IJMPD..13.1549G}
F.~{Govoni}, L.~{Feretti}, {Magnetic Fields in Clusters of Galaxies},
  International Journal of Modern Physics D 13 (2004) 1549--1594.

\bibitem{Wolfe:1992ab}
A.~M. Wolfe, K.~M. Lanzetta, A.~L. Oren, Magnetic fields in damped ly-alpha
  systems, Astrophys. J. 388 (1992) 17--22.

\bibitem{Clarke:2000bz}
T.~E. Clarke, P.~P. Kronberg, H.~Boehringer, A new radio - x-ray probe of
  galaxy cluster magnetic fields, Astrophys. J. 547 (2001) L111--L114.

\bibitem{Xu:2005rb}
Y.~Xu, P.~P. Kronberg, S.~Habib, Q.~W. Dufton, A faraday rotation search for
  magnetic fields in large scale structure, Astrophys. J. 637 (2006) 19--26.

\bibitem{2011arXiv1112.0340B}
D.~J. {Barnes}, D.~{Kawata}, K.~{Wu}, {Cosmological Simulations using GCMHD+},
  ArXiv e-prints\href {http://arxiv.org/abs/1112.0340}
  {\path{arXiv:1112.0340}}.

\bibitem{Liddle:2000booka}
A.~R. Liddle, D.~H. Lyth, Cosmological Inflation and Large-Scale Structure,
  Cambridge University Press, 2000.

\bibitem{Lemoine:1995dm}
D.~Lemoine, M.~Lemoine, Primordial magnetic fields in string cosmology, Phys.
  Rev. D 52~(4) (1995) 1955--1962.

\bibitem{Quashnock:1989jm}
J.~M. Quashnock, A.~Loeb, D.~N. Spergel, Magnetic field generation during the
  cosmological qcd phase transition, Astrophys. J. Lett. 344 (1989) L49--L51.

\bibitem{1996PhRvD..53..662B}
G.~{Baym}, D.~{B{\"o}deker}, L.~{McLerran}, {Magnetic fields produced by phase
  transition bubbles in the electroweak phase transition}, Phys. Rev. D 53
  (1996) 662--667.

\bibitem{Caprini:2001nb}
C.~Caprini, R.~Durrer, Gravitational wave production: A strong constraint on
  primordial magnetic fields, Phys. Rev. D 65 (2001) 023517.

\bibitem{2000PhR...331..283M}
M.~{Maggiore}, {Gravitational wave experiments and early universe cosmology},
  Physics Reports 331 (2000) 283--367.
\newblock \href {http://arxiv.org/abs/arXiv:gr-qc/9909001}
  {\path{arXiv:arXiv:gr-qc/9909001}}, \href
  {http://dx.doi.org/10.1016/S0370-1573(99)00102-7}
  {\path{doi:10.1016/S0370-1573(99)00102-7}}.

\bibitem{2005APh....23..313C}
R.~H. {Cyburt}, B.~D. {Fields}, K.~A. {Olive}, E.~{Skillman}, {New BBN limits
  on physics beyond the standard model from \^{}4He}, Astroparticle Physics 23
  (2005) 313--323.
\newblock \href {http://arxiv.org/abs/arXiv:astro-ph/0408033}
  {\path{arXiv:arXiv:astro-ph/0408033}}, \href
  {http://dx.doi.org/10.1016/j.astropartphys.2005.01.005}
  {\path{doi:10.1016/j.astropartphys.2005.01.005}}.

\bibitem{2007ApJS..170..288H}
G.~{Hinshaw}, M.~R. {Nolta}, C.~L. {Bennett}, R.~{Bean}, O.~{Dor{\'e}}, M.~R.
  {Greason}, M.~{Halpern}, R.~S. {Hill}, N.~{Jarosik}, A.~{Kogut},
  E.~{Komatsu}, M.~{Limon}, N.~{Odegard}, S.~S. {Meyer}, L.~{Page}, H.~V.
  {Peiris}, D.~N. {Spergel}, G.~S. {Tucker}, L.~{Verde}, J.~L. {Weiland},
  E.~{Wollack}, E.~L. {Wright}, {Three-Year Wilkinson Microwave Anisotropy
  Probe (WMAP) Observations: Temperature Analysis}, Astrophys. J. Suppl 170
  (2007) 288--334.

\bibitem{2007ApJS..170..335P}
L.~{Page}, G.~{Hinshaw}, E.~{Komatsu}, M.~R. {Nolta}, D.~N. {Spergel}, C.~L.
  {Bennett}, C.~{Barnes}, R.~{Bean}, O.~{Dor{\'e}}, J.~{Dunkley}, M.~{Halpern},
  R.~S. {Hill}, N.~{Jarosik}, A.~{Kogut}, M.~{Limon}, S.~S. {Meyer},
  N.~{Odegard}, H.~V. {Peiris}, G.~S. {Tucker}, L.~{Verde}, J.~L. {Weiland},
  E.~{Wollack}, E.~L. {Wright}, {Three-Year Wilkinson Microwave Anisotropy
  Probe (WMAP) Observations: Polarization Analysis}, Astrophys. J. Suppl 170
  (2007) 335--376.

\bibitem{2007ApJS..170..377S}
D.~N. {Spergel}, R.~{Bean}, O.~{Dor{\'e}}, M.~R. {Nolta}, C.~L. {Bennett},
  J.~{Dunkley}, G.~{Hinshaw}, N.~{Jarosik}, E.~{Komatsu}, L.~{Page}, H.~V.
  {Peiris}, L.~{Verde}, M.~{Halpern}, R.~S. {Hill}, A.~{Kogut}, M.~{Limon},
  S.~S. {Meyer}, N.~{Odegard}, G.~S. {Tucker}, J.~L. {Weiland}, E.~{Wollack},
  E.~L. {Wright}, {Three-Year Wilkinson Microwave Anisotropy Probe (WMAP)
  Observations: Implications for Cosmology}, Astrophys. J. Suppl 170 (2007)
  377--408.

\bibitem{2009ApJS..180..330K}
E.~{Komatsu}, J.~{Dunkley}, M.~R. {Nolta}, C.~L. {Bennett}, B.~{Gold},
  G.~{Hinshaw}, N.~{Jarosik}, D.~{Larson}, M.~{Limon}, L.~{Page}, D.~N.
  {Spergel}, M.~{Halpern}, R.~S. {Hill}, A.~{Kogut}, S.~S. {Meyer}, G.~S.
  {Tucker}, J.~L. {Weiland}, E.~{Wollack}, E.~L. {Wright}, {Five-Year Wilkinson
  Microwave Anisotropy Probe Observations: Cosmological Interpretation},
  Astrophys. J. Suppl 180 (2009) 330--376.

\bibitem{Ross:2008ze}
A.~J. {Ross}, R.~J. {Brunner}, A.~D. {Myers}, {Normalization of the Matter
  Power Spectrum via Higher Order Angular Correlations of Luminous Red
  Galaxies}, Astrophys. J. 682 (2008) 737--744.

\bibitem{Rozo:2007yt}
E.~Rozo, et~al., {Cosmological Constraints from SDSS maxBCG Cluster
  Abundances}\href {http://arxiv.org/abs/astro-ph/0703571}
  {\path{arXiv:astro-ph/0703571}}.

\bibitem{2011ApJS..192...18K}
E.~{Komatsu}, K.~M. {Smith}, J.~{Dunkley}, C.~L. {Bennett}, B.~{Gold},
  G.~{Hinshaw}, N.~{Jarosik}, D.~{Larson}, M.~R. {Nolta}, L.~{Page}, D.~N.
  {Spergel}, M.~{Halpern}, R.~S. {Hill}, A.~{Kogut}, M.~{Limon}, S.~S. {Meyer},
  N.~{Odegard}, G.~S. {Tucker}, J.~L. {Weiland}, E.~{Wollack}, E.~L. {Wright},
  {Seven-year Wilkinson Microwave Anisotropy Probe (WMAP) Observations:
  Cosmological Interpretation}, Astrophys. J. Suppl 192 (2011) 18--+.
\newblock \href {http://arxiv.org/abs/1001.4538} {\path{arXiv:1001.4538}},
  \href {http://dx.doi.org/10.1088/0067-0049/192/2/18}
  {\path{doi:10.1088/0067-0049/192/2/18}}.

\bibitem{2010PhRvD..81b3002W}
S.~{Wang}, {New primordial-magnetic-field limit from the latest LIGO S5 data},
  Phys. Rev. D 81~(2) (2010) 023002.

\bibitem{LIGO_S5}
{The LIGO Scientific Collaboration and The Virgo Collaboration}, An upper limit
  on the stochastic gravitational-wave background of cosmological origin,
  Nature 460 (2009) 990.

\bibitem{2010PhRvD..82h3005K}
T.~{Kahniashvili}, A.~G. {Tevzadze}, S.~K. {Sethi}, K.~{Pandey}, B.~{Ratra},
  {Primordial magnetic field limits from cosmological data}, Phys. Rev. D
  82~(8) (2010) 083005.

\bibitem{2011PhRvD..83l3533P}
D.~{Paoletti}, F.~{Finelli}, {CMB constraints on a stochastic background of
  primordial magnetic fields}, Phys. Rev.~(12) (2011) 123533--+.
\newblock \href {http://arxiv.org/abs/1005.0148} {\path{arXiv:1005.0148}},
  \href {http://dx.doi.org/10.1103/PhysRevD.83.123533}
  {\path{doi:10.1103/PhysRevD.83.123533}}.

\bibitem{1968ApJ...151..459S}
J.~{Silk}, {Cosmic Black-Body Radiation and Galaxy Formation}, Astrophys. J.
  151 (1968) 459.

\bibitem{Seshadri:2000ky}
T.~R. Seshadri, K.~Subramanian, {CMBR Polarization Signals from Tangled
  Magnetic Fields}, Phys. Rev. Lett. 87 (2001) 101301.

\bibitem{1979ZhPmR..30..719S}
A.~A. Starobinskii, ZhETF Pis ma Redaktsiiu 30 (1979) 719.

\bibitem{1982PhLB..115..189R}
V.~A. Rubakov, M.~V. Sazhin, A.~V. Veryaskin, Physics Letters B 115 (1982) 189.

\bibitem{1985SvA....29..607P}
A.~G. Polnarev, Soviet Astronomy 29 (1985) 607.

\bibitem{Pritchard:2004qp}
J.~R. Pritchard, M.~Kamionkowski, Ann. Phys. 318 (2005) 2.

\bibitem{1980ARA&A..18..537S}
R.~A. {Sunyaev}, I.~B. {Zeldovich}, {Microwave background radiation as a probe
  of the contemporary structure and history of the universe}, ARA. and A. 18
  (1980) 537--560.
\newblock \href {http://dx.doi.org/10.1146/annurev.aa.18.090180.002541}
  {\path{doi:10.1146/annurev.aa.18.090180.002541}}.

\bibitem{cosmomc}
A.~Lewis, S.~Bridle, Cosmological parameters from cmb and other data: a monte-
  carlo approach, Phys. Rev. D 66 (2002) 103511.

\bibitem{Peebles:1980booka}
P.~J.~E. Peebles, The Large-Scale Structure of the Universe, Princeton
  University Press, 1980.

\bibitem{Yamamoto:1997qc}
K.~Yamamoto, N.~Sugiyama, H.~Sato, Evolution of small-scale cosmological baryon
  perturbations and matter transfer functions, Astrophys. J. 501 (1997) 442.

\bibitem{Cole:2005sx}
S.~Cole, et~al., {The 2dF Galaxy Redshift Survey: Power-spectrum analysis of
  the final dataset and cosmological implications}, Mon. Not. Roy. Astron. Soc.
  362 (2005) 505--534.

\bibitem{Tegmark:2006az}
M.~Tegmark, et~al., {Cosmological Constraints from the SDSS Luminous Red
  Galaxies}, Phys. Rev. D74 (2006) 123507.

\bibitem{Jedamzik:1996wp}
K.~Jedamzik, V.~Katalinic, A.~V. Olinto, Damping of cosmic magnetic fields,
  Phys. Rev. D 57 (1998) 3264--3284.

\bibitem{Subramanian:1997gi}
K.~Subramanian, J.~D. Barrow, Magnetohydrodynamics in the early universe and
  the damping of noninear alfven waves, Phys. Rev. D 58 (1998) 083502.

\bibitem{Banerjee:2004df}
R.~Banerjee, K.~Jedamzik, The evolution of cosmic magnetic fields: From the
  very early universe, to recombination, to the present, Phys. Rev. D 70 (2004)
  123003.

\bibitem{Dunkley:2008ie}
J.~Dunkley, et~al., {Five-Year Wilkinson Microwave Anisotropy Probe (WMAP)
  Observations: Likelihoods and Parameters from the WMAP data}, Astrophys. J.
  Suppl. 180 (2009) 306--329.

\bibitem{Finelli:2008xh}
F.~Finelli, F.~Paci, D.~Paoletti, {The Impact of Stochastic Primordial Magnetic
  Fields on the Scalar Contribution to Cosmic Microwave Background
  Anisotropies}, Phys. Rev. D78 (2008) 023510.
\newblock \href {http://arxiv.org/abs/0803.1246} {\path{arXiv:0803.1246}},
  \href {http://dx.doi.org/10.1103/PhysRevD.78.023510}
  {\path{doi:10.1103/PhysRevD.78.023510}}.

\bibitem{2010arXiv1006.4242S}
J.~R. {Shaw}, A.~{Lewis}, {Constraining Primordial Magnetism}, ArXiv
  e-prints\href {http://arxiv.org/abs/1006.4242} {\path{arXiv:1006.4242}}.

\bibitem{Hinshaw:2008kr}
G.~{Hinshaw}, J.~L. {Weiland}, R.~S. {Hill}, N.~{Odegard}, D.~{Larson}, C.~L.
  {Bennett}, J.~{Dunkley}, B.~{Gold}, M.~R. {Greason}, N.~{Jarosik},
  E.~{Komatsu}, M.~R. {Nolta}, L.~{Page}, D.~N. {Spergel}, E.~{Wollack},
  M.~{Halpern}, A.~{Kogut}, M.~{Limon}, S.~S. {Meyer}, G.~S. {Tucker}, E.~L.
  {Wright}, {Five-Year Wilkinson Microwave Anisotropy Probe Observations: Data
  Processing, Sky Maps, and Basic Results}, Astrophys. J 180 (2009) 225--245.

\bibitem{Kuo:2006ya}
C.~L. Kuo, P.~A.~R. Ade, J.~J. Bock, J.~R. Bond, C.~R. Contaldi, M.~D. Daub,
  J.~H. Goldstein, W.~L. Holzapfel, A.~E. Lange, M.~Lueker, M.~Newcomb, J.~B.
  Peterson, C.~Reichardt, J.~Ruhl, M.~C. Runyan, Z.~Staniszweski, Improved
  measurements of the cmb power spectrum with acbar, Astrophys. J. 664~(2)
  (2007) 687--701.

\bibitem{Sievers:2005gj}
J.~L. {Sievers}, C.~{Achermann}, J.~R. {Bond}, L.~{Bronfman}, R.~{Bustos},
  C.~R. {Contaldi}, C.~{Dickinson}, P.~G. {Ferreira}, M.~E. {Jones}, A.~M.
  {Lewis}, B.~S. {Mason}, J.~{May}, S.~T. {Myers}, N.~{Oyarce}, S.~{Padin},
  T.~J. {Pearson}, M.~{Pospieszalski}, A.~C.~S. {Readhead}, R.~{Reeves}, A.~C.
  {Taylor}, S.~{Torres}, Implications of the cosmic background imager
  polarization data, Astrophys. J. 660 (2007) 976--987.

\bibitem{Jones:2005yb}
W.~C. Jones, et~al., A measurement of the angular power spectrum of the cmb
  temperature anisotropy from the 2003 flight of boomerang, Astrophys. J. 647
  (2006) 823--832.

\bibitem{1980PhRvL..45.1980B}
J.~R. {Bond}, G.~{Efstathiou}, J.~{Silk}, {Massive neutrinos and the
  large-scale structure of the universe}, Physical Review Letters 45 (1980)
  1980--1984.
\newblock \href {http://dx.doi.org/10.1103/PhysRevLett.45.1980.2}
  {\path{doi:10.1103/PhysRevLett.45.1980.2}}.

\bibitem{Lesgourgues:2006nd}
J.~Lesgourgues, S.~Pastor, {Massive neutrinos and cosmology}, Phys. Rept. 429
  (2006) 307--379.

\bibitem{DeBernardis:2008qq}
F.~De~Bernardis, P.~Serra, A.~Cooray, A.~Melchiorri, {An improved limit on the
  neutrino mass with CMB and redshift-dependent halo bias-mass relations from
  SDSS, DEEP2, and Lyman-Break Galaxies}, Phys. Rev. D78 (2008) 083535.

\bibitem{Ichiki:2008rh}
K.~Ichiki, Y.-Y. Keum, {Neutrino Masses from Cosmological Probes in Interacting
  Neutrino Dark-Energy Models}, JHEP 06 (2008) 058.

\bibitem{Ichiki:2008ye}
K.~Ichiki, M.~Takada, T.~Takahashi, {Constraints on Neutrino Masses from Weak
  Lensing}\href {http://arxiv.org/abs/0810.4921} {\path{arXiv:0810.4921}}.

\bibitem{2008RPPh...71h6201O}
E.~W. {Otten}, C.~{Weinheimer}, {Neutrino mass limit from tritium {$\beta$}
  decay}, Reports on Progress in Physics 71~(8) (2008) 086201--+.
\newblock \href {http://dx.doi.org/10.1088/0034-4885/71/8/086201}
  {\path{doi:10.1088/0034-4885/71/8/086201}}.

\bibitem{Elgaroy:2002bi}
O.~Elgaroy, et~al., {A new limit on the total neutrino mass from the 2dF galaxy
  redshift survey}, Phys. Rev. Lett. 89 (2002) 061301.

\bibitem{2003ApJS..148..119K}
E.~{Komatsu}, A.~{Kogut}, M.~R. {Nolta}, C.~L. {Bennett}, M.~{Halpern},
  G.~{Hinshaw}, N.~{Jarosik}, M.~{Limon}, S.~S. {Meyer}, L.~{Page}, D.~N.
  {Spergel}, G.~S. {Tucker}, L.~{Verde}, E.~{Wollack}, E.~L. {Wright},
  {First-Year Wilkinson Microwave Anisotropy Probe (WMAP) Observations: Tests
  of Gaussianity}, Astrophys. J. Suppl 148 (2003) 119--134.

\bibitem{2011PhRvD..83l3003S}
M.~{Shiraishi}, D.~{Nitta}, S.~{Yokoyama}, K.~{Ichiki}, K.~{Takahashi}, {Cosmic
  microwave background bispectrum of tensor passive modes induced from
  primordial magnetic fields}, Phys. Rev. D 83~(12) (2011) 123003.

\bibitem{2010PhRvD..82l3006T}
P.~{Trivedi}, K.~{Subramanian}, T.~R. {Seshadri}, {Primordial magnetic field
  limits from cosmic microwave background bispectrum of magnetic passive scalar
  modes}, Phys. Rev. D 82~(12) (2010) 123006.

\bibitem{2009JCAP...06..021C}
C.~{Caprini}, F.~{Finelli}, D.~{Paoletti}, A.~{Riotto}, {The cosmic microwave
  background temperature bispectrum from scalar perturbations induced by
  primordial magnetic fields}, JCAP 6 (2009) 21.

\bibitem{2009PhRvL.103h1303S}
T.~R. {Seshadri}, K.~{Subramanian}, {Cosmic Microwave Background Bispectrum
  from Primordial Magnetic Fields on Large Angular Scales}, Physical Review
  Letters 103~(8) (2009) 081303.

\bibitem{2005PhRvD..72f3002B}
I.~{Brown}, R.~{Crittenden}, {Non-Gaussianity from cosmic magnetic fields},
  Phys. Rev. D 72~(6) (2005) 063002.

\bibitem{2006MNRAS.372.1060T}
H.~{Tashiro}, N.~{Sugiyama}, {Probing primordial magnetic fields with the 21-cm
  fluctuations}, Mon. Not. Roy. Astron. Soc. 372 (2006) 1060--1068.

\end{thebibliography}







\end{document}